\documentclass{aa}  
\usepackage{amssymb}
\usepackage{graphicx}
\usepackage[authoryear]{natbib}
\usepackage[figuresright]{rotating}

\usepackage{txfonts}
\newcommand{\teff}{\ifmmode T_{\rm eff} \else T$_{\mathrm{eff}}$\fi}
\newcommand{\logg}{\ifmmode \log g \else $\log g$\fi}
\newcommand{\lL}{\ifmmode \log \frac{L}{L_{\odot}} \else $\log \frac{L}{L_{\odot}}$\fi}
\newcommand{\mdot}{$\dot{M}$}
\newcommand{\myr}{M$_{\odot}$ yr$^{-1}$}
\newcommand{\vsini}{$V$ sin$i$}
\newcommand{\vinf}{$v_{\infty}$}

\newcommand{\kms}{km s$^{-1}$}
\newcommand{\msun}{\ifmmode M_{\odot} \else M$_{\odot}$\fi}
\newcommand{\zsun}{\ifmmode Z_{\odot} \else Z$_{\odot}$\fi}
\newcommand{\lsun}{\ifmmode L_{\odot} \else L$_{\odot}$\fi}
\newcommand{\rsun}{\ifmmode R_{\odot} \else R$_{\odot}$\fi}
\newcommand{\qh}{\ifmmode Q_{\rm H} \else $Q_{\rm H}$\fi}
\newcommand{\qhei}{\ifmmode Q_{\ion{He}{i}} \else $Q_{\ion{He}{i}}$\fi}
\newcommand{\ha}{H$_{\alpha}$}
\newcommand{\mum}{\ifmmode \mu m \else $\mu m$\fi}

\begin{document}
   \title{Properties of WNh stars in the Small Magellanic Cloud: \hspace{2cm} evidence for homogeneous evolution. \thanks{Based on observations collected at the ESO Very Large Telescope (programs 077.D-0029, 071.D-0272, 074.D-0696)}}

   \subtitle{}

   \author{F. Martins\inst{1}
          \and
          D.J. Hillier\inst{2}
          \and
          J.C. Bouret\inst{3}
          \and
          E. Depagne\inst{4}
          \and
          C.Foellmi\inst{5}
          \and
          S. Marchenko\inst{6}
          \and
          A.F. Moffat\inst{7}
          }

   \offprints{F. Martins}

   \institute{GRAAL--CNRS, Universit\'e Montpellier II -- UMR 5024, Place Eug\`ene Bataillon, F--34095, Montpellier Cedex 05, France \\
              \email{martins AT graal.univ-montp2.fr}
         \and 
             Department of Physics and Astronomy, University of Pittsburgh, 3941 O'Hara street, Pittsburgh, PA 15260, USA\\
         \and
             LAM -- CNRS, Universit\'e de Provence -- UMR 6110, 38 rue Fr\'ed\'eric Joliot-Curie, F--13388, Marseille Cedex 13, France\\
         \and
             Las Cumbres Observatory, 6740 Cortona Dr, suite 102, 93117 Goleta, CA, USA\\
         \and
             LAOG -- CNRS, Observatoire de Grenoble, Universit\'e Joseph Fourier -- UMR 5571, BP53, F-38041, Grenoble Cedex 9, France\\
	 \and
             Department of Physics and Astronomy, Western Kentucky University, Bowling Green, KY, USA\\
         \and
             D\'epartement de Physique, Universit\'e de Montr\'eal, succursalle Centre-ville, Montr\'eal, QC H3C 3J7, Canada\\
             }

   \date{Received 22 September 2008; accepted 13 November 2008}

\authorrunning{F. Martins et al.}
\titlerunning{WNh stars in the SMC}

% \abstract{}{}{}{}{} 
% 5 {} token are mandatory
 
  \abstract
   {}
   {To assess the nature of three WNh stars in the SMC, to constrain stellar evolution beyond the main sequence at low metallicity, and to investigate the metallicity dependence of the clumping properties of massive stars. }
   {We compute atmosphere models to derive the stellar and wind properties of the three WNh targets. A FUV/UV/optical/near-infrared analysis gives access to temperatures, luminosities, mass loss rates, terminal velocities and stellar abundances.}
   {All stars still have a large hydrogen mass fraction in their atmosphere, and show clear signs of CNO processing in their surface abundances. One of the targets can be accounted for by normal stellar evolution. It is a star with initial mass around 40-50 \msun\ in, or close to, the core He burning phase. The other two objects must follow a peculiar evolution, governed by fast rotation. In particular, one object is likely evolving homogeneously due to its position blue-ward of the main sequence and its high H mass fraction. The clumping factor of one star is found to be 0.15$\pm$0.05. This is comparable to values found for Galactic Wolf-Rayet stars, indicating that within the uncertainties, the clumping factor does not seem to depend on metallicity. }
   {}


   \maketitle

%%#####################################################################
%%-------------------------------   Introduction  --------------------- 

\section{Introduction}
\label{s_intro}

Massive stars have gained considerable interest in the last years in various fields of astrophysics. In addition to being well recognized actors in Galactic astronomy (HII regions, bubbles, ISM dynamics and winds, triggered star formation, chemical enrichment) their role in extragalactic physics has considerably grown. Massive stars are known to govern starburst events locally and at high redshift \citep[e.g.][]{steidel96}, being responsible for outflows \citep{veilleux05} and intense episodes of star formation in luminous infrared galaxies \citep{genzel98}. The association of some GRBs with core collapse supernova \citep{galama98,hjorth03} make fast rotating massive stars the likely progenitors of GRBs \citep{ww93}. However, the conditions under which a massive star will actually end up in such an event are still poorly known. Fast rotation is of course a key parameter, but the way the star loses angular momentum, and thus its mass loss history, is equally important. With mass loss decreasing at lower metallicity, GRBs are thought to occur in low Z environments. Observations of GRB hosts tend to confirm this expectation \citep{proch04}. However, the existence of an upper metallicity limit for the formation of GRBs is not established. Massive stars are also thought to be responsible for the re-ionization of the Universe at very high redshift (e.g. Schaerer at al.\ 2002 and references therein). Indeed, the first (population III) stars may have been very massive objects with very high effective temperatures due to their metal free composition \citep{bromm99,nu99}. 

In spite of this growing importance of massive stars, their evolution is not fully understood even in the local Universe. It is not clear at present which phases a star of a given initial mass will go through during its life. For example, the initial conditions for a star to enter the red supergiant or luminous blue variable stages are rather blurry. Similarly, the evolutionary connections between various types of Wolf-Rayet stars is poorly constrained. These difficulties result from two facts. The first one is that massive star evolution is governed by mass loss, and our knowledge of mass loss rates through the upper part of the HR diagram is still patchy. The second problem is due to rotation: stars with similar initial masses but different rotational velocities will follow different paths in the HR diagram, and thus will evolve through different states \citep{mm00a}. These two ingredients (mass loss and rotation) are the keys to massive star evolution. An additional complication occurs because both of the above factors depend on metallicity: stars lose less mass at low metallicity \citep{puls00}, and at the same time they rotate faster \citep{martayan07}. Consequently, it is not only crucial to know the mass loss and rotation of all types of massive stars, but it is also important to constrain their metallicity dependence. Recent results indicate a \mdot\ $\sim Z^{0.83}$ behavior for O stars \citep{mokiem07c}. Theoretical predictions and observational constraints also exist for various types of Wolf-Rayet stars \citep{vink05,goetz08,paul02}. A rather clear picture of the {\it relative} strength of stellar winds at different metallicities is thus emerging. But the question of their {\it absolute} strength is not settled. The reason is clumping.

There is now no doubt that massive--star winds are not homogeneous. Evidence comes from both indirect and direct tracers. The former comprise electron scattering wings of emission lines \citep{hil91}, too--weak observed UV line profiles \citep{paul02b,hil03,massa03,jc03,jc05,fullerton06}, shapes of X-ray lines \citep{oskinova04}, submm excess in continuum emission \citep{blomme02}, linear--polarization continuum variations \citep{stlouis93}. Direct indications are the observations of global moving substructures on emission--line profiles \citep{eversb98,lm99,lm08}. Clumping affects mass--loss determination of massive stars because it changes the density, which in turn modifies the shape of wind lines. In practice, clumping is usually quantified by means of a filling factor $f$. For a given $f$, recombination lines will be fitted, to first order, with a mass loss rate approximately $1/\sqrt{f}$ times lower compared to a homogeneous atmosphere model. Knowledge of the clumping properties of massive--star winds is thus a key to the quantitative understanding of massive star evolution.

Studies of Galactic stars indicate $f$ factors ranging from 0.01 to 0.2 \citep{hk98,hil01,jc05,fullerton06}. Such factors may not be constant throughout the wind of a massive star \citep{puls06}. In addition, if inhomogeneities are due to line driving instabilities \citep{runacres02}, a metallicity dependence can be expected. It is however not constrained at present. A first step in this direction was made by \citet{mar07} (hereafter paper I). In this study, the dynamical properties of clumps directly observed in emission lines of three SMC WR stars were found to be very similar to those of Galactic objects.

In the present study, we analyze the three stars observed in paper I by means of atmosphere models. We constrain the stellar and wind properties, which help us understand their evolutionary status. Being located in the SMC, these stars give us a clearer picture of stellar evolution beyond the main sequence in a low metallicity environment. We find that evolution at very high rotational rate is most likely at work, as suspected for the formation of GRBs. In addition, one can pinpoint the value of the clumping factor $f$ for one star, showing no systematic difference with values found for Galactic stars.

In the following, we briefly review the observational material in Sect.\ \ref{s_spec} before explaining the modelling technique in Sect.\ \ref{s_models}. We then present the results for each star in Sect.\ \ref{s_results} and discuss them in Sect.\ \ref{s_disc}. Sect.\ \ref{s_conc} summarizes our conclusions.

%%#####################################################################
%%-------------------------------   Observations  --------------------- 

\section{Spectroscopic data}
\label{s_spec}

The three stars analyzed here (SMC-WR1, SMC-WR2, SMC-WR4) were observed with the UV-Visual Echelle Spectrograph (UVES) on VLT/UT2 at ESO. Observations were conducted on August 27-28$^{th}$ 2006. The regions between 3930--6030 was targeted. A slit width of 1.5\arcsec\ was used and observations were conducted under an airmass varying between 1.5 and 2.7, and a seeing fluctuating between 0.8 and 2\arcsec. The data were reduced using the ESO UVES Pipeline. The reduction process 
includes bias and inter-order background subtraction, for both the objects 
frames and the flat-fields calibration images, optimal extraction  of the 
objects (that includes removing skylines and cosmics rejection), division by 
a flat-field extracted with the same weighed profile as the object, 
wavelength calibration and merging of all overlapping orders after a 
rebinning to a constant wavelength. No spectrophotometric standard was observed, so our spectra are not flux calibrated. The final spectra have a resolving power of $\sim$ 20000 and a signal to noise ratio between 120 and 160. More details about the observations and data reduction can be found in \cite{mar07}. The spectra we use in the following are the combination of all spectra obtained during the observing run.

We have also used the weighted mean spectra covering the \ha\ region presented by \citet{foel03a}. They have a low resolution (R$\sim$800) and result from observations with various telescopes and spectrographs \citep[see details in Table 2 of][]{foel03a}. These spectra are not flux calibrated.

In addition to the optical spectra, we used archival IUE data for stars SMC-WR2 and SMC-WR4 (spectra SWP7026 and a weighted mean -- according to exposure time -- of SWP06195 and SWP10727 respectively). They were taken in the low resolution mode (R $\sim$ 250) of the instrument. Two high resolution spectra exist for SMC-WR4, but their S/N is low and they clearly suffer from reduction artifacts (e.g. oscillating continuum). For those two stars we also retrieved FUSE archival data (spectra G036101000 and G03900301000). They have a resolving power of about 20000.  

Finally, ESO--SOFI spectra of the sample stars covering the J and H band (also K--band for SMC-WR4) were used. The data on SMC-WR4 were presented by \citet{paul00}. For SMC-WR1 and SMC-WR2, we used unpublished data kindly provided by Paul Crowther. They were obtained in November 2003 (only SMC-WR1) and November 2004 (SMC-WR1 and SMC-WR2) with a 0.6\arcsec\ slit. Their resolution is $\sim$ 500 and they are not flux calibrated. They were reduced following the same procedure as described in \citet{paul00}.

%%#####################################################################
%%-------------------------------   Modelling   ----------------------- 

\section{Modelling}
\label{s_models}

The stellar and wind properties of the stars analyzed here have been derived using non-LTE, line blanketed models including winds. The models have been computed with the code CMFGEN. A detailed description of the code is given in \cite{hm98}. Here, we recall only a few points relevant for the understanding of the present study.

\textit{Line-blanketing:} a super-level approximation is used to allow the inclusion of a large number of atomic levels from various elements. In our models, H, He, C, N, O, Ne, P, Si, S, Fe, Ni have been included. In total, the populations of $\sim$1150 levels and $\sim$3500 super-levels were calculated through the resolution of the statistical equilibrium equations. A constant microturbulent velocity (20\kms) was adopted for the calculation of the atmospheric structure. For the formal solution of the transfer equation leading to the emergent spectrum, a microturbulent velocity ranging from 10--50 \kms\ (depending on the star) to 0.1$\times$\vinf\ from bottom to top of the atmosphere was chosen. Stellar abundances were chosen to be 1/5$^{th}$ solar \citep{gs98}.

\textit{Clumping:} CMFGEN allows the inclusion of clumping with the following formalism. A volume filling factor $f$ is assumed to vary monotonically with depth, starting from a value of 1 at the photosphere and reaching a maximum value $f_{\infty}$ at the outer boundary of the atmosphere. In practice, $f$ is parametrized as a function of the local wind velocity $v$

\begin{center}
\begin{equation}
f = f_{\infty} + (1-f_{\infty})e^{-\frac{v}{v_{cl}}}
\label{eq_clump}
\end{equation}  
\end{center}

\noindent where $v_{cl}$ is a parameter fixing the velocity at which clumping starts to become significant. In practice, we chose $v_{cl}$ = 100 \kms\ so that clumping starts above the sonic point. Lower values have been reported for O stars \citep{jc05} but are difficult to explain theoretically \citep{runacres02}. \cite{puls06} showed that in O stars, clumping varied non monotonically: after an increase ($f$ decreases) up to the region around the sonic point, clumping decreases in the outer wind. Theoretical simulations by \cite{runacres02} tend to predict a similar trend. However, this behavior has not been constrained for Wolf-Rayet stars. Hence, we adopt a conservative approach and keep on working with a monotonic clumping law, keeping in mind that it might not be fully realistic.

\textit{Density structure:} Until recently a major problem with the analysis of W-R stars was that the density structure had to be specified using an assumed velocity law.  However work by \citet{goetz05,goetz08}  have allowed velocity laws (and mass loss rates) to be derived for some W-R stars using a formulation of radiation driving applicable to optically thick winds. A significant limitation of the work is that the volume filling factor $f$, and its variation with radius, must be assumed since as we have seen clumping properties are still poorly constrained.

Unfortunately, CMFGEN is not yet able to self-consistently compute the velocity law. It still assumes some analytical expression above (approximately) the sonic point: the so-called $\beta$ velocity law is connected to the photospheric structure. Below the sonic point, a photospheric structure is either parametrized by a scale height formalism \citep{hm98} or is taken from alternative atmosphere models, usually TLUSTY \citep{lh03}. In the present analysis, we have improved this formalism. We now require that hydrostatic equilibrium is satisfied below the sonic point. In practice, we proceeded as follows. We first started with a photospheric TLUSTY structure connected to a $\beta$ velocity law. With this structure, we started to converge the models. After a few iterations (usually 10-15), the predicted level populations were used to compute the radiative acceleration. This acceleration was subsequently injected in the hydrostatic equilibrium equation from which an altered density structure was derived at depth, below the sonic point. This ensures a higher degree of consistency between the hydrodynamic and atmospheric structures. Above V(sonic)/2, we adopted the usual beta velocity law formulation (see below). With this revised structure, the models were converged for another 10-15 iterations before another modification of the photospheric structure was performed. We ran 3--4 of these structure iterations before converging the atmosphere models. 
In practice, the main inconsistency in the final density structure (lack of radiative acceleration) was found around the sonic point. This reflects the fact that we are specifying the wind velocity law rather than deriving it, that the critical point is close to the sonic point, and that the disagreement depends on the clumping law adopted. \\

With these models, we have constrained the main stellar and wind parameters using the following procedure:

\begin{itemize}

\item[$\bullet$] \textit{Temperature}. As usual for Wolf-Rayet stars, we have derived both an effective temperature representing the radiation flux at a Rosseland optical depth of 2/3, and a temperature T$_{*}$ at $\tau$=20. Both values can be significantly different in the case of a very dense wind for which $\tau$=2/3 is reached in a region where the atmosphere is no longer static. In practice, the temperature of the model was adjusted to correctly reproduce the \textit{relative} strength of \ion{N}{iii} $\lambda\lambda$4635,4641, \ion{N}{iii} $\lambda\lambda$4510,4515, \ion{N}{iv} $\lambda$4058, \ion{N}{v} $\lambda\lambda$4604, 4620 and \ion{N}{v} $\lambda$4944 (when present). Infrared He lines -- \ion{He}{i} $\lambda$1.08 \mum, \ion{He}{ii} $\lambda$1.17 \mum -- were also used and usually gave consistent results (see Sect.\ \ref{s_results}) . In addition for one star (SMC-WR4) we could use the classical \ion{He}{i}/\ion{He}{ii} optical line ratio since the optical \ion{He}{i} lines were strong enough. We estimate the uncertainty on the derived temperatures to be of the order of 3000 K for SMC-WR2 and SMC-WR4, 7000 K for SMC-WR1 (for which \ion{N}{iii} is not present and \ion{N}{iv} is weak).

\item[$\bullet$] \textit{Luminosity and extinction}. For stars SMC-WR2 and SMC-WR4, the luminosity and local extinction was determined from the fit of the SED. The IUE (flux calibrated) spectra have been used in conjunction with optical photometry (our optical spectra are not flux calibrated) to adjust the stellar luminosity and extinction. Table \ref{tabphotom} summarizes the photometric data for each star. FUSE data were not used since 1) the extinction laws in the far-UV are not well constrained and 2) the IUE and FUSE spectra did not show the same flux level in the region around 1200 \AA\ (especially for SMC-WR2), pointing to an uncertainty in the flux calibration (either for FUSE or IUE data). Given point 1, we relied preferentially on IUE spectra. Infrared data were not used since the extinction law in the near-IR is not constrained for the SMC. The Galactic (SMC) extinctions laws of \citet{seaton79} and \citet{howarth83} \citep{bouchet85} were used. The SMC extinction is stronger than the Galactic one in the UV (they are similar in the optical), so that using both IUE and optical data helped constrain the local, SMC extinction and the foreground, Galactic E(B-V). A distance modulus of 19.0 (corresponding to a distance of 63.2 kpc) was assumed. This value was reported by \citet{diben97} and is intermediate between low values (18.80) and high (19.20) values reported in the literature \citep{cole98,caputo99,groen00,harries03,hilditch05,keller08}. For star SMC-WR1, in the absence of (F)UV data, we only used the absolute V magnitude. 
From the observed B-V and the theoretical value expected for hot massive stars \citep[(B-V)$_{0}$=-0.28, see][]{mp06} we derive E(B-V)=0.24 which, together with an assumed distance modulus of 19.0, implies M$_{V}$=-4.57. We adjusted the luminosity to reproduce this value.

\item[$\bullet$] \textit{Abundances}. The ratio of He to H was constrained from the relative strength of the optical \ion{He}{ii} lines (especially those at 4686 and 5412 \AA) to the Balmer lines (mainly H$_{\beta}$ and  H$_{\delta}$). The carbon abundance was obtained from \ion{C}{iv} $\lambda\lambda$5801, 5812 and, to a lesser extent, from \ion{C}{iv} $\lambda\lambda$1548, 1551. The nitrogen content was derived from the same lines used to constrain the temperature. Note that the latter depends on the \textit{relative} strength of those lines, while the N abundance affects the \textit{absolute} strength of the lines. In addition, \ion{N}{iv} $\lambda$1486 and \ion{N}{iv} $\lambda$1720 were also use as a secondary indicator. \ion{O}{v} $\lambda$5597 and \ion{O}{v} $\lambda$5604 are used to provide upper limits on the oxygen content.

\item[$\bullet$] \textit{Mass loss rate and terminal velocity}. \mdot\ was derived from the numerous emission and P-Cygni lines present throughout the (F)UV and optical spectrum. We used the blue-ward velocity edge of the absorption part of P-Cygni profiles to derive \vinf.

 \item[$\bullet$] \textit{Velocity field}. The slope of the velocity field (described by the so-called $\beta$ parameter) was derived from the shape of the \ion{He}{ii} 4686 emission profile. This profile reacts to a combination of change of density due to a different velocity field and to the subsequent change of formation depth of the line implied by the modified atmospheric density structure. In practice, a narrower, more peaked profile results from a larger $\beta$. This is somewhat different from the usual behavior encountered among O stars. The difference is due to the higher wind density affecting the line formation depth in WR stars, while for O stars it is not significantly changed. 

\item[$\bullet$] \textit{Clumping}. The clumping factor $f_{\infty}$ was constrained from the strength of the electron scattering wing of \ion{He}{ii} $\lambda$4686. As shown by \cite{hil91}, this part of the spectrum is especially sensitive to the degree of homogeneity in the star's atmosphere: more structured winds lead to weaker electron scattering wings. This method was already used by \cite{paul00} for the study of the star SMC-WR4.

\end{itemize}

Finally, in the FUSE range we have added HI and H$_{2}$ interstellar absorptions on top of our synthetic spectrum. We have not attempted a detailed fit of these interstellar lines. Instead, we have chosen column densities reproducing the main absorption lines.

%%#####################################################################
%%-------------------------------   Results   ----------------------- 

\section{Results}
\label{s_results}

The results of our analysis in terms of derived parameters are summarized in Table \ref{tab_param}. In the following, we briefly comment on the analysis of each individual star.

\begin{figure*}
\begin{center}
\begin{minipage}[b]{0.4\linewidth} % A minipage that covers half the page
\centering
\includegraphics[width=7cm]{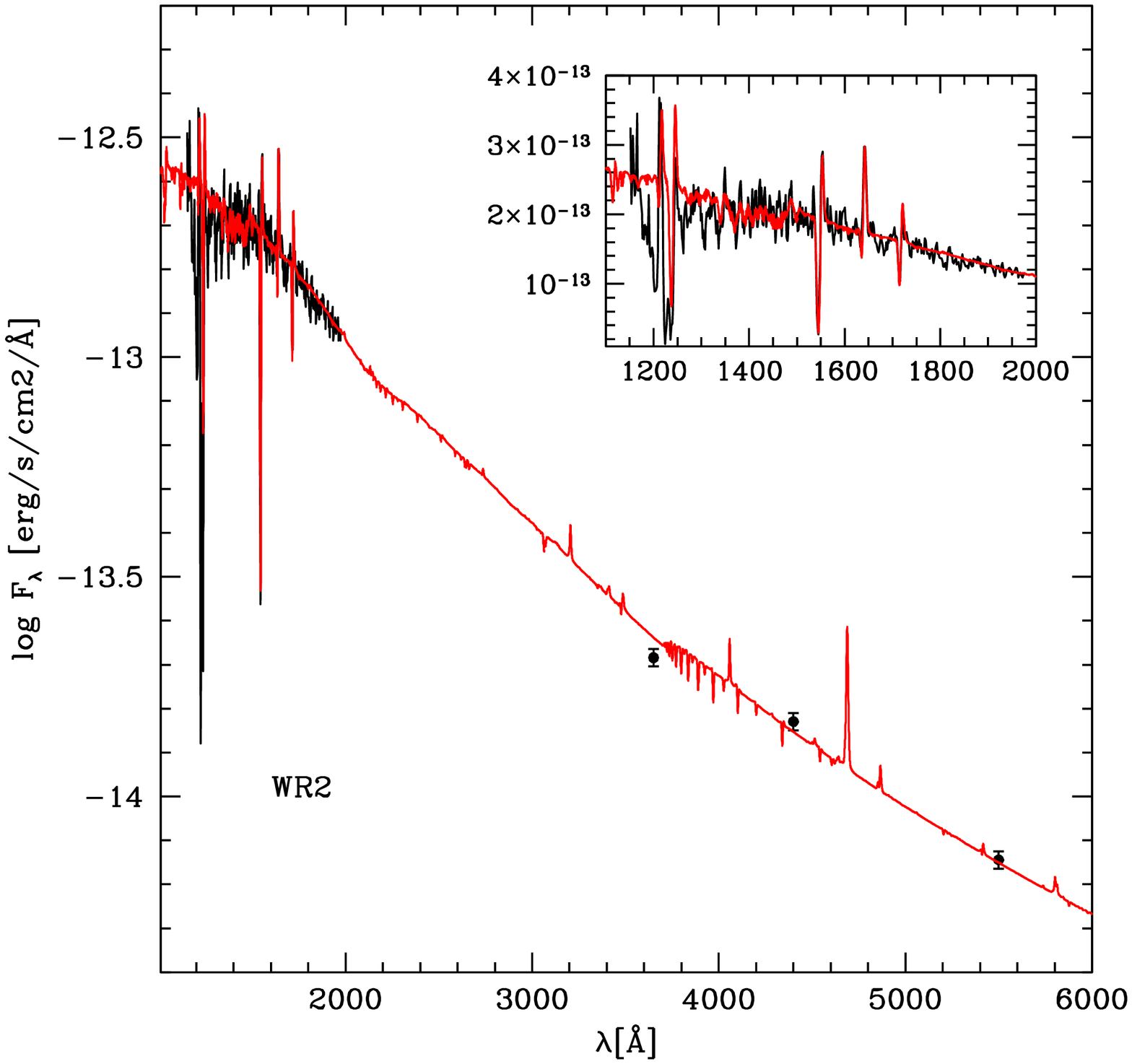}
%\caption{}
\end{minipage}
\hspace{0.5cm} % To get a little bit of space between the figures
\begin{minipage}[b]{0.4\linewidth}
\centering
\includegraphics[width=7cm]{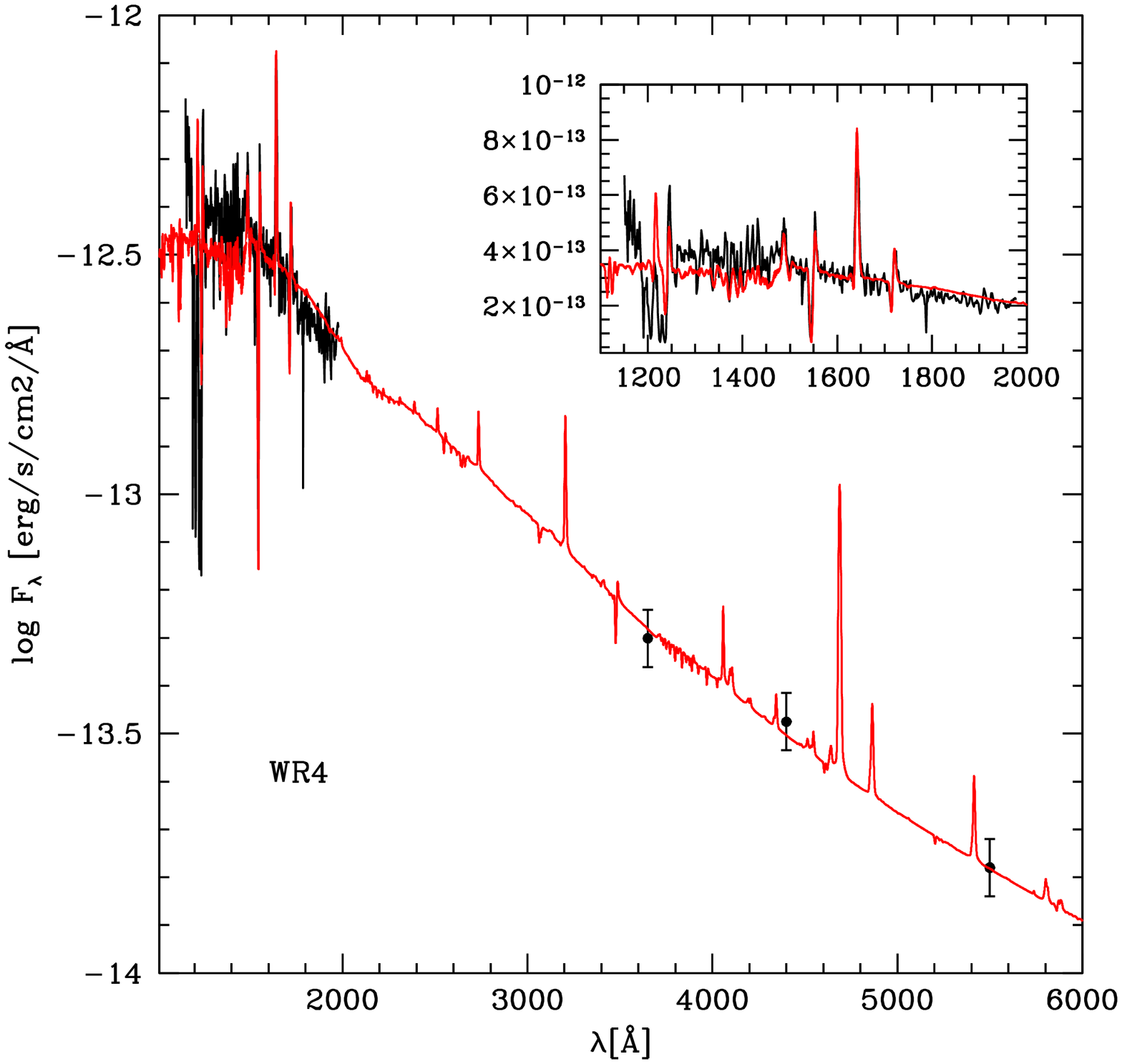}
%\caption{En liten bild till}
\end{minipage}
\caption{SED fitting for SMC-WR2 (left panel) and SMC-WR4 (right panel). The IUE spectra are reported as well as the fluxes in the UBV bands. E(B-V)=0.03+0.08 (Galactic + SMC) and 0.04+0.10 are required for SMC-WR2 and SMC-WR4 respectively. A distance of 63.2kpc is assumed \citep{diben97}. This SED fitting yields the luminosities: $10^{5.50} L_{\odot}$ for SMC-WR2, $10^{5.90} L_{\odot}$ for SMC-WR4.} \label{fitsed}
\end{center}
\end{figure*}

\begin{table}
\renewcommand{\thefootnote}{\thempfootnote}
\centering
\caption{Photometry of the target stars from \citet{md01} together with the derived extinction.} \label{tabphotom}
\begin{minipage}{0.5\textwidth}
\centering
\begin{tabular}{clrrrr}
\hline
Star       & ST  &  U & B & V & E(B-V) \\
           &     &    &   &   & Gal.+ SMC \\
\hline       
SMC-WR1    & WN3h & 14.22 & 15.13 & 15.17 & 0.24 \\
SMC-WR2    & WN5h & 13.26 & 14.08 & 14.26 & 0.03+0.06 \\
SMC-WR4    & WN6h & 12.48 & 13.25 & 13.37 & 0.04+0.10\\
\hline
\end{tabular}
\end{minipage}
\end{table}

%%-------------------------------   Results WR4   --------------------- 

\subsection{SMC-WR4}
\label{s_res_wr4}
\begin{figure*}
\begin{center}
\begin{minipage}[b]{0.4\linewidth} % A minipage that covers half the page
\centering
\includegraphics[width=7cm]{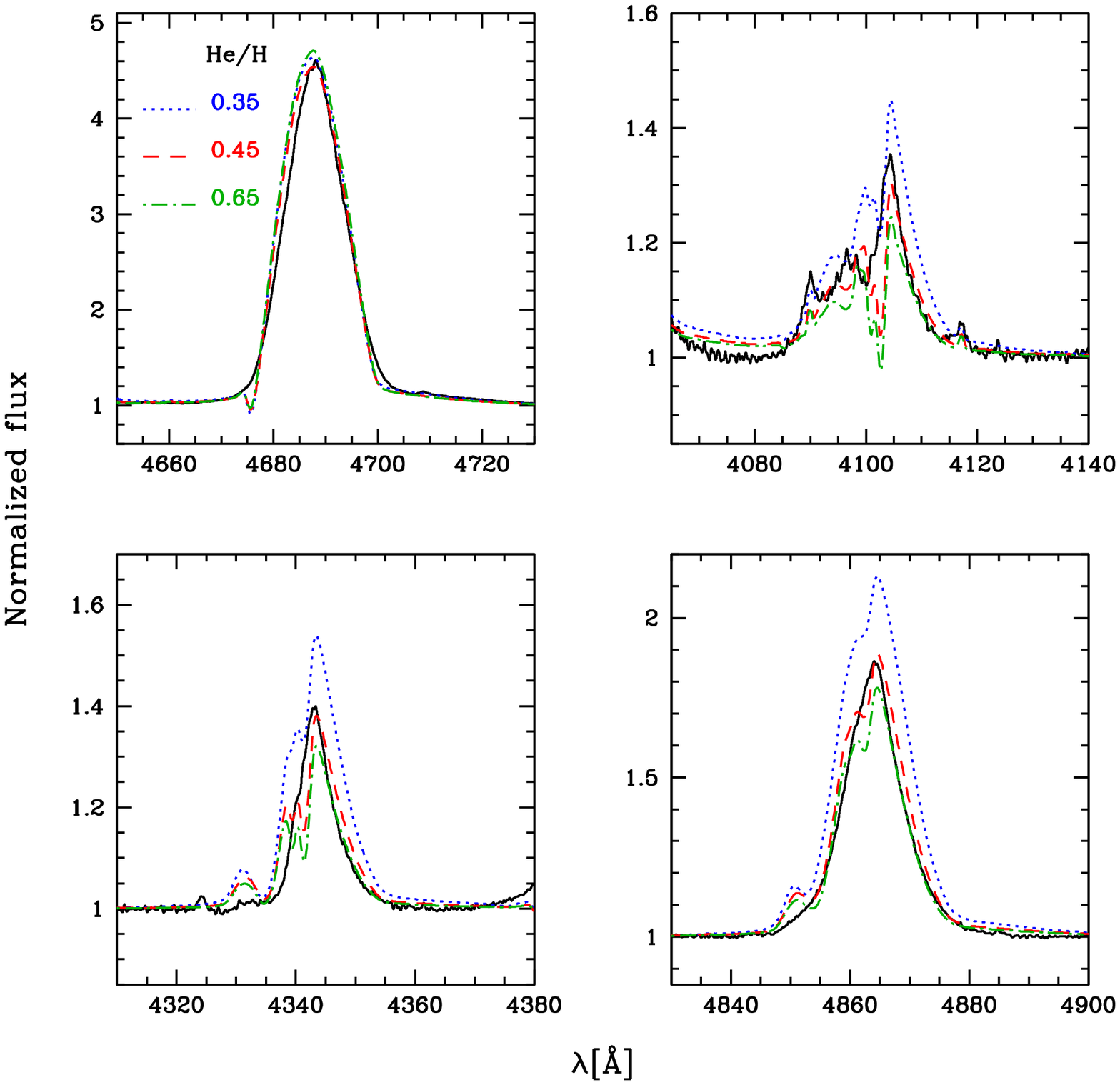}
%\caption{}
\end{minipage}
\hspace{0.5cm} % To get a little bit of space between the figures
\begin{minipage}[b]{0.4\linewidth}
\centering
\includegraphics[width=7cm]{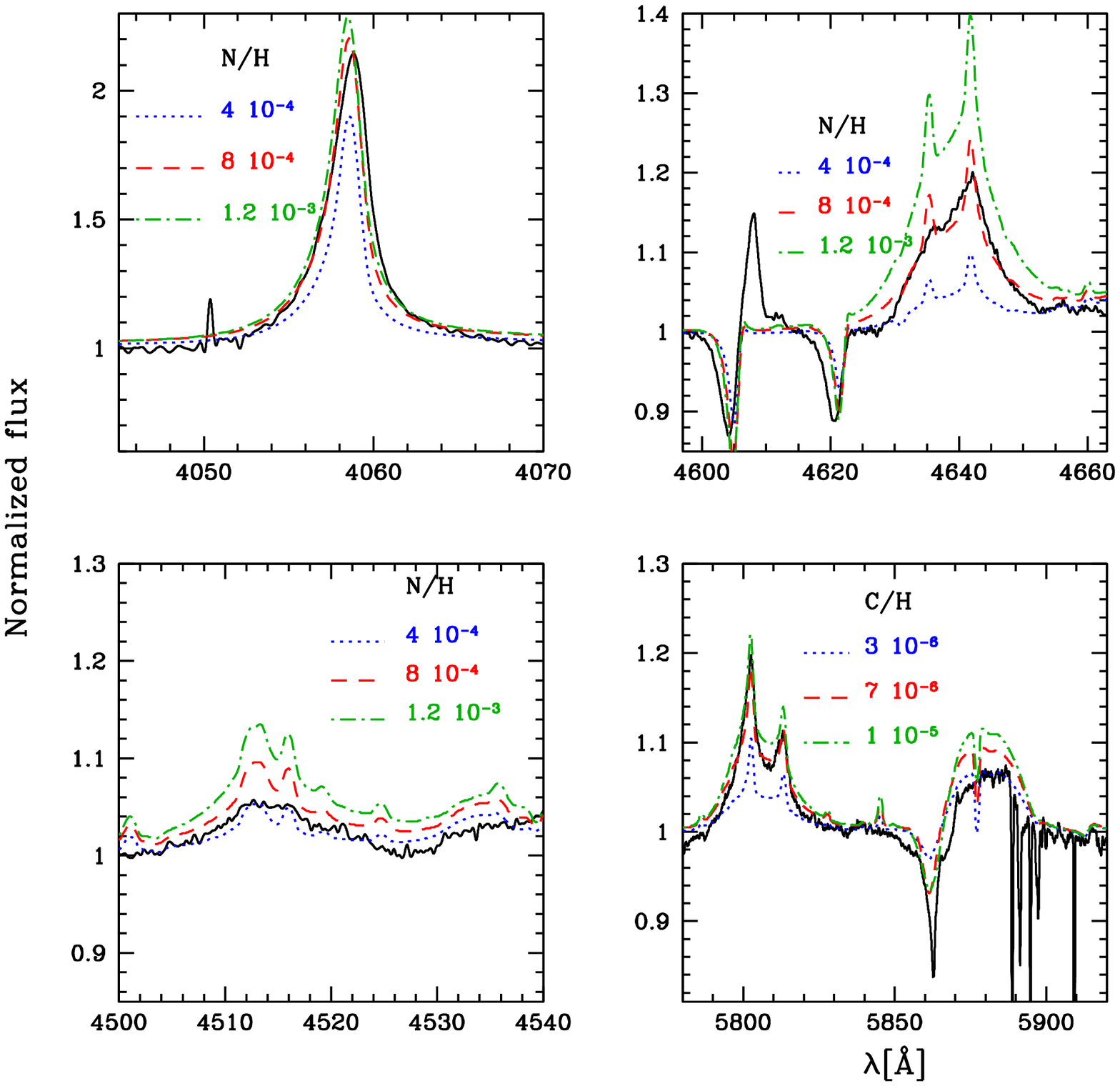}
%\caption{En liten bild till}
\end{minipage}
\caption{He (left) and CN (right) abundance determination for SMC-WR4. The black solid line is the observed spectrum, the interrupted colored lines are models with different He/H (C/H, N/H) ratios.}\label{ab_wr4}
\end{center}
\end{figure*}

The fit of the SED is shown in Fig.\ \ref{fitsed}. The error bars on the optical data reflect the dispersion of measured V magnitudes found in the literature. A luminosity of $10^{5.90} L_{\odot}$ is derived. The best fit of the normalized spectra of SMC-WR4 is shown in Fig.\ \ref{fit_wr4}. The only major difficulty concerns the emission line red-ward of \ion{N}{v} $\lambda$4604, which is not reproduced. But overall, the fit is of excellent quality. An effective temperature of 42000 K gives the best fit of the optical N lines. Slightly lower values ($\sim$ 41000 K) are preferred for the near-IR He lines (especially \ion{He}{i} $\lambda$1.08 \mum), still consistent with the N lines results within the uncertainties. This difference can be explained if the star's spectrum shows variability. Indeed, the near-IR and optical data have been obtained at different epochs. Since the star shows line profile variability (see paper I), simultaneous optical/infrared observations are needed to investigate this discrepancy. The simplified approach to treat clumping might also affect the detailed line profiles. For the FUV interstellar absorption, we found that HI (resp. H$_{2}$) column densities of $10^{22.0} cm^{-2}$ (resp. $10^{19.7}$) gave reasonable results. Individual abundances for He, C, N and O (upper limit) were derived from our analysis. Their determination is illustrated in Figs.\ \ref{ab_wr4}. A medium degree of He enrichment is detected in SMC-WR4, together with a rather strong N overabundance and a carbon depletion. The typical uncertainties on these abundance determinations are of the order 30 to 50$\%$, as seen from Fig.\ \ref{ab_wr4} \footnote{These errors do not include any systematics due to uncertainties in the atomic data}. Finally the mass loss rate determination gives a value of $8 \times\ 10^{-6}$ \msun. Strong optical emission lines would require slightly larger values of the order  $10^{-5}$ \msun, while near-IR H and He lines tend to indicate slightly lower values ( $\sim 7 \times\ 10^{-6}$ \msun). \footnote{Note that our best fit model in Fig. \ref{fit_wr4} has a slightly smaller mass loss rate (0.1 dex) than the best fit model of Fig. \ref{ab_wr4} in order to best match both optical/UV and near-infrared lines. In Fig.\ \ref{ab_wr4}, \mdot\ is scaled to fit \ion{He}{ii} $\lambda$4686}.

\begin{figure*}
\begin{center}
\begin{minipage}[b]{0.4\linewidth} % A minipage that covers half the page
\centering
\includegraphics[width=7cm]{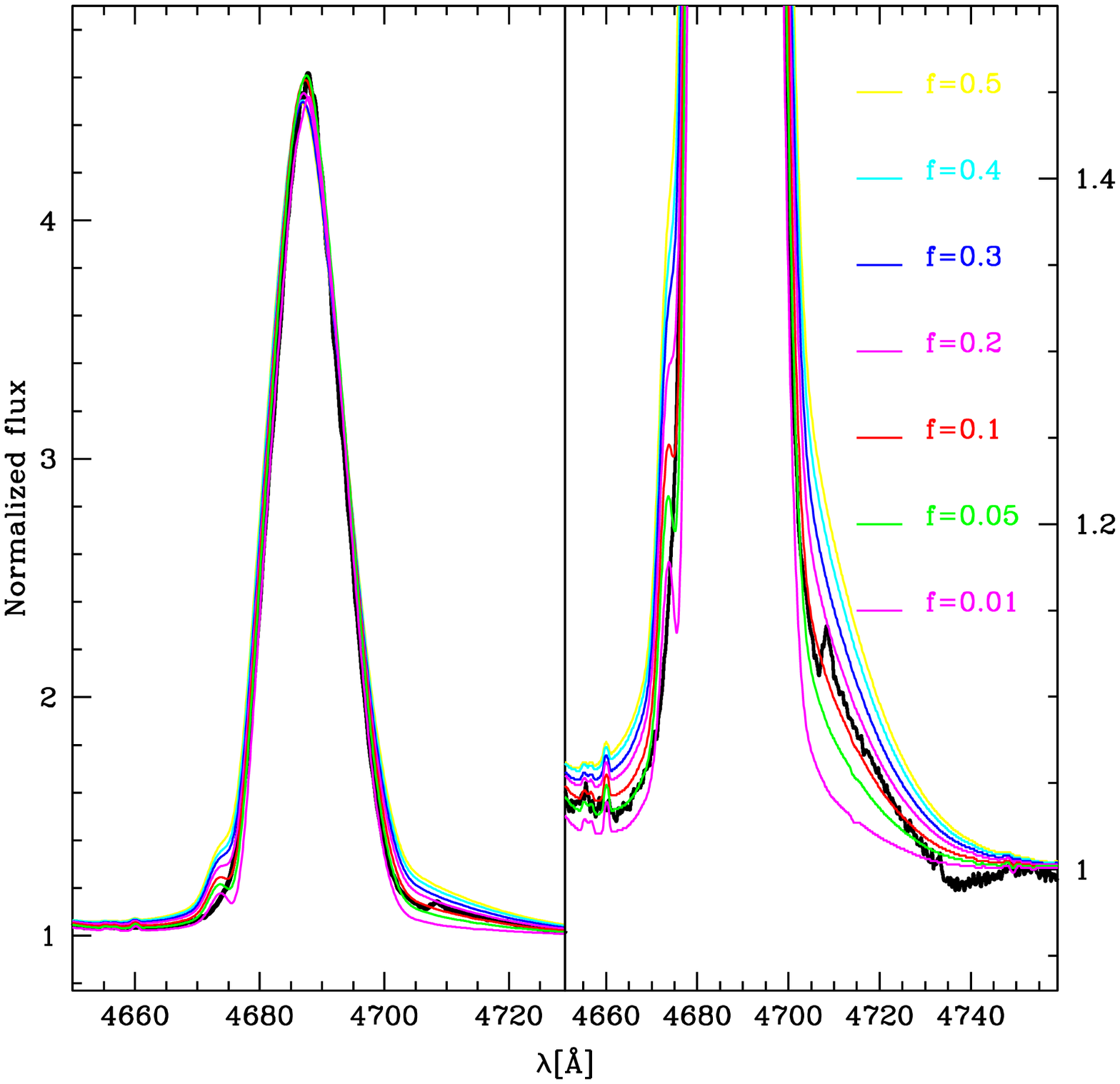}
%\caption{}
\end{minipage}
\hspace{0.5cm} % To get a little bit of space between the figures
\begin{minipage}[b]{0.4\linewidth}
\centering
\includegraphics[width=7cm]{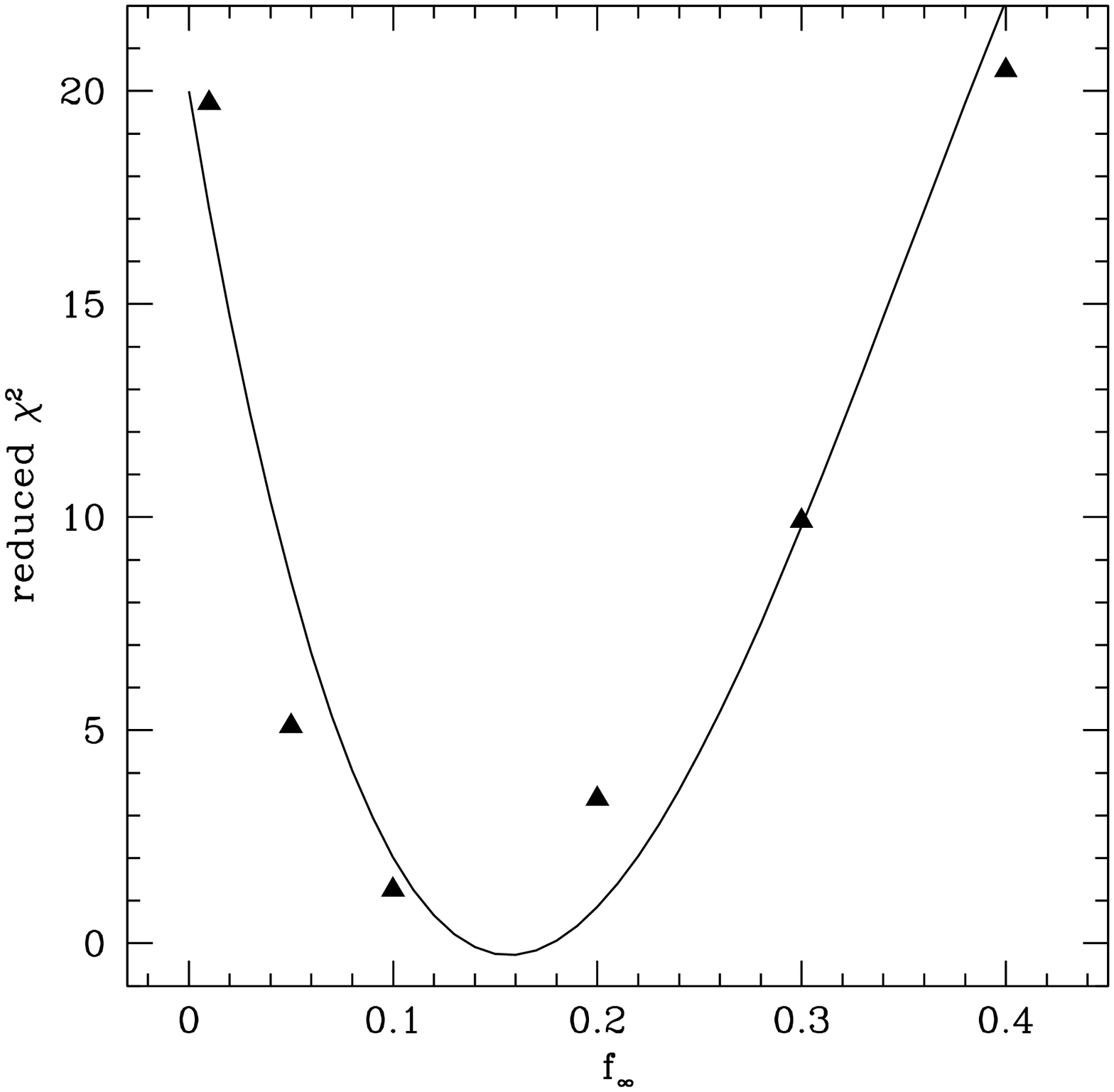}
%\caption{En liten bild till}
\end{minipage}
\caption{Determination of clumping parameter $f_{\infty}$. {\it Left}: Effect of the clumping parameter f$_{\infty}$ on the electron scattering wing of HeII 4686 for star SMC-WR4. {\it Right}: Quantification of the goodness of the fit of the red wing of HeII 4686. The solid line is a simple third order polynomial fit. A value of 0.15$\pm$0.05 is preferred for the clumping factor $f_{\infty}$.} \label{f_effect_wr4}
\end{center}
\end{figure*}

The left panel of Fig.\ref{f_effect_wr4} shows various models with different clumping parameters ($f_{\infty}$ ranging from 0.01 to 0.5). For each model, the mass--loss rate is scaled to properly account for the emission of \ion{He}{ii} $\lambda$4686. One clearly sees that the strength of the red electron scattering wing of the line depends on the amount of structure in the wind. From this feature, we see that a model with $f_{\infty} \sim$  0.1 best represents the observed profile. For a more quantitative statement, we have estimated the goodness of the fit of this red wing by means of a $\chi^2$ analysis of the region between 4702 and 4730 \AA. The results are shown in Fig.\ \ref{f_effect_wr4}, right panel. A value of 0.15$\pm$0.05 is preferred. 

SMC-WR4 was studied by \citet{paul00} in the optical and near infrared, using CMFGEN. The parameters we obtain are very close to his. The only notable difference is the slightly larger luminosity we derive, although within the error bars the agreement is acceptable. The difference comes from the different distances to the SMC that were adopted: we use 63.2 kpc while \citet{paul00} assumes a distance of 60.2 kpc. Using his distance, we would derive a luminosity of $10^{5.8} L_{\odot}$, in good agreement with Crowther's result. We also note that Crowther's fit also showed the problem reported for the mass loss rate determination: near-IR lines tend to indicate lower \mdot\ than optical lines.

%%-------------------------------   Results SMC-WR2   --------------------- 

\subsection{SMC-WR2}
\label{s_res_wr2}

The fit of the SED of SMC-WR2 is shown in Fig.\ \ref{fitsed}. The error bars on the optical data reflect the dispersion of measured V magnitudes. We derive a luminosity of $10^{5.50} L_{\odot}$. Fig.\ \ref{fit_wr2} shows our best fit of the normalized spectra. The overall quality of the fit is as good as for SMC-WR4. HI and H$_{2}$ interstellar absorption lines have been added in the FUSE range with the following column densities: $10^{21.7} cm^{-2}$ for HI and $10^{19.3} cm^{-2}$ for H$_{2}$. The optical N lines indicate an effective temperature of $\sim$ 42500 K. We note that with this \teff\ the \ion{He}{i} $\lambda$1.08 \mum\ line is underpredicted. Obtaining a stronger emission requires a reduction of \teff\ by $\sim$ 5000 K. But in that case the optical spectrum is not reproduced anymore: the \ion{N}{v} lines vanish completely, \ion{N}{iv} $\lambda$4058 \AA\ is strongly reduced and the \ion{N}{iii} lines become much too strong. The same explanation as for SMC-WR4 could be invoked. Clearly, coordinated multiwavelength observations (as soon possible with instruments such as VLT/X--Shooter) are necessary to solve this issue. One can also note that the S/N ratio of the near-IR data is lower than for the optical data, increasing the uncertainty on the temperature determination from NIR lines.  Abundance determinations were performed as for SMC-WR4. Only a small He enrichment is required to fit the He and H lines of the spectrum. However, evidence for CNO processing comes from the existence of a C depletion and a N overabundance. We note that both optical and near-IR emission lines are well reproduced by the a single value of the mass loss rate.

Unlike in SMC-WR4, \ion{He}{ii} $\lambda$4686 is a rather weak emission line. Consequently, its red wing is more sensitive to details of the data reduction and normalization. We have attempted to derive the clumping factor $f_{\infty}$ as for SMC-WR4, but it turned out that no conclusions could be drawn due to these observational uncertainties. In particular, the exact position of the continuum is uncertain due to residuals of the flat field correction and to imperfect order merging. Other classical clumping indicators are found in the FUV--UV range. \ion{O}{v} $\lambda$1371 and \ion{N}{iv} $\lambda$1720 have been shown to depend on the degree of inhomogeneities \citep[e.g.][]{jc05}. However, the IUE spectra of SMC-WR2 have a too low spectral resolution and these lines cannot be used. In the FUSE range, \ion{P}{v} $\lambda\lambda$1118, 1128 are known to depend on the amount of clumping. \citet{fullerton06} showed that very low filling factors were necessary to account for the strength of this doublet, while \citet{paul02} and \citet{hil03} obtained satisfactory fits with $f_{\infty} \sim 0.1$. \citet{fullerton06} highlighted the high sensitivity of this doublet to the ionization structure, which can be influenced by shocks and X--rays. This effect is probably reduced in dense WR star winds  compared to O stars, but might still be present. Besides, \citet{oskinova07} revealed that depending on which type of clumping was present (micro vs macro--clumping) the line profiles had different shapes. Finally, \citet{paul02} showed that the phosphorus abundance strongly influenced the shape of this unsaturated line. Given these multiple dependencies, we refrained from using \ion{P}{v} $\lambda\lambda$1118, 1128 as a clumping diagnostic in the present study. We simply adopted $f_{\infty}$=0.1 for SMC-WR2.

%%-------------------------------   Results SMC-WR1   --------------------- 

\subsection{SMC-WR1}
\label{s_res_wr1}

Unfortunately, only optical and near-IR spectra are available for SMC-WR1. The analysis is thus more limited than for the two other stars. In particular, in the absence of UV P-Cygni lines, we adopted a terminal velocity of 1800 \kms\ which turned out to give an excellent fit of the optical emission lines. The uncertainty on \teff\ (and consequently on L) is larger for this star since only a limited number of lines from successive ionization states is available. In particular, although several \ion{N}{v} lines are used, the only \ion{N}{iv} line (at 4058 \AA) is very weak. It can be used to place a lower limit on \teff. The upper limit comes from the shape of the \ion{N}{v} $\lambda\lambda$4604, 4620 lines which become narrower and weaker at too high effective temperature. In practice, the uncertainty on \teff\ is $\pm$ 7000 K. This translates into an uncertainty on \lL\ of 0.15 dex. In spite of these increased uncertainties, we were able to derive a number of stellar and wind parameters. Fig.\ \ref{fit_wr1} shows our best fit. Overall, the quality is good: optical and near-IR lines can be reproduced with a single set of stellar and wind parameters. The \ion{He}{ii} lines of the Brackett series are slightly too deep and narrow. A change of $\beta$ does not help since it degrades the fit of \ion{He}{ii} $\lambda$4686. Similarly, increasing \vinf\ broadens too much the other emission lines. \ion{He}{ii} $\lambda$5412 is slightly underestimated. Improving its fit by increasing \mdot, \teff\ or He/H alters the fits of all other lines. Presumably, remaining uncertainties in the density structure are responsible for this discrepancy.

\begin{figure}[]
\centering
\includegraphics[width=9cm]{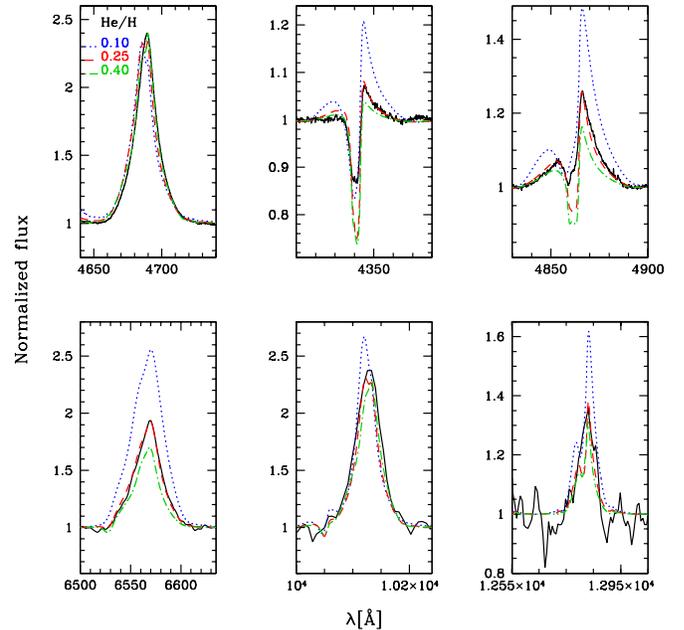}
\caption{Determination of the He to H ratio for star SMC-WR1. The observed spectrum is the black solid line. The mass loss rates of the models with different He/H ratios were adjusted to give a correct fit of the \ion{He}{ii} $\lambda$4686 \AA\ emission line. A value of He/H=0.25 is preferred.}\label{He_effect_wr1}
\end{figure}

Like SMC-WR2 and SMC-WR4, SMC-WR1 is slightly He enriched and strongly N rich. Fig.\ \ref{He_effect_wr1} illustrates our He/H determination. As we will see later, this quantity is especially important to understand the evolution of SMC-WR1. From the weakness of \ion{C}{iv} and \ion{O}{v} lines in the red optical spectrum, one can put upper limits on the C and O content (see Table \ref{tab_param}), indicating depletion of these two elements. As for SMC-WR2, \ion{He}{ii} $\lambda$4686 is not strong enough to allow a correct clumping factor determination. We adopted $f_{\infty}$=0.1.

\begin{table*}
\renewcommand{\thefootnote}{\thempfootnote}
\centering
\caption{Parameters of our best fit models. Uncertainties are: 3kK (7kK for SMC-WR1) on temperatures, 0.1 dex on luminosities (0.15 for SMC-WR1), 0.2 dex on \mdot, 100 \kms\ on \vinf, 0.05 on He/H, 30 to 50$\%$ on abundances. } \label{tab_param}
\begin{minipage}{\textwidth}

\centering
\begin{tabular}{clrrrrrrrrrrrrr}
\hline
Star & ST  & T$_{*}$ & \teff  & \lL  & R$_{*}$ & $\log \dot{M}$ & f$_{\infty}$ & $\beta$ & \vinf   & He/H & X   & X(C)   & X(N) & X(O)\\
     &     &  [kK]   &  [kK] &      & R$_{\odot}$ & [\myr]         &            &        & [\kms]  &  \#  & (\%) &(\%)   & (\%) & (\%)\\
\hline                                                                
SMC-WR1    & WN3h &  65.4   &  60.5 & 5.75 & 5.8 & -5.55          &  0.10\footnote{Adopted}       & 1.0 & 1800    & 0.25 &    0.50 & $<$0.004 & 0.12 & $<$0.04 \\
SMC-WR2    & WN5h &  44.8   &  42.5 & 5.50 & 9.3 & -5.85          &  0.10\footnotemark[\value{mpfootnote}]       & 1.0 & 1000    & 0.15 &    0.62 & 0.005    & 0.35 & -- \\
SMC-WR4    & WN6h &  43.1   &  42.0 & 5.90 & 16.0& -5.00          &  0.15\footnote{Derived from $\chi^{2}$ fit, mass loss rate scaled from model closest to minimum (f$_{\infty}$=0.1).}       & 1.0 & 1000    & 0.45 &    0.35 & 0.003    & 0.40 & $<$0.15 \\
\hline
\end{tabular}
\end{minipage}
\end{table*}

%%#####################################################################
%%-------------------------------   Discussion   ----------------------- 

\section{Discussion}
\label{s_disc}

In the next sections, we first comment on our results on the clumping factor of SMC-WR4 in the context of a metallicity dependence of clumping properties. We then move to the interpretation of the evolutionary status of the three WNh stars analyzed in the present study.

%%-------------------------------   Wind & clumping   --------------------- 

\subsection{Winds and clumping}
\label{s_wind}

In paper I, a thorough investigation of moving features in strong emission lines revealed that the dynamical properties of clumps in the SMC were very similar to those of Galactic WR stars. Here, we have attempted to constrain the clumping factor at low Z. For star SMC-WR4, we have carried out a detailed study of the clumping factor. We have quantitatively shown that the preferred value for $f_{\infty}$ (see Eq.\ \ref{eq_clump}) was 0.15$\pm$0.05. The $\chi^2$ curve of Fig.\ \ref{f_effect_wr4} unambiguously confirms that lower and larger values are excluded. 

Studies of Galactic WR stars have been the first to reveal that clumping was present in massive-star winds \citep[e.g.][]{moffat88,stlouis89}. Quantitative determinations of clumping factors were carried out by various groups using different atmosphere codes. \citet{hk98} analyzed three WN (one WN5, one WN6 and one WN8) stars and one WC star and found that a clumping filling--factor of 0.25 was best suited for WN stars, although 0.06 was still compatible within the uncertainties. The latter value was best for the WC star. Similarly, \citet{schmutz97} found that $f_{\infty}$ = 0.25 led to a good fit of the emission lines of the WN4 star HD50896. \citet{hm99} found that $f_{\infty} = 0.1$ was best suited to fit the UV-optical-IR spectrum of a Galactic WC5 star. \citet{hil01} carried out an extensive study of the spectra of $\eta$ Car and found that a clumping factor of $\sim$0.1 better fitted the electron scattering wings of both H and HeI lines compared to $\sim$0.2-0.3. \citet{herald01} analyzed two Galactic WN8 stars and concluded that $0.05 < f < 0.2$ from the same diagnostics. \citet{paul02,paul06} analyzed WC stars in the LMC and the Galaxy and found that f$_{\infty}$=0.1 was preferred compared to 1.0 and 0.01. However, no other intermediate values were tested. \citet{paco04} found $f_{\infty}$ values between 0.08 and 0.15 for five WNLh stars in the Arches cluster. Hence, a value of 0.1--0.2 seems to be common among Galactic WR stars and related, post main sequence objects.  

For SMC-WR4, we have shown that $f_{\infty}$ was definitely between 0.10 and 0.20 (see Fig.\ref{f_effect_wr4}). Keeping in mind that we are dealing with a single measurement for the SMC, it looks as if there is no difference between the low metallicity environment of the SMC and the Galaxy in terms of clumping factor. Or more accurately, if there is any variation, it is not larger than $\sim$ 0.1, which is the uncertainty of current determinations. This is consistent with the results of paper I. 

From a theoretical point of view, the hydrodynamical simulations of \citet{owo88}, \citet{runacres02} and \citet{runacres05} show that instabilities grow at a rate proportional to $g_{rad}$, the line radiative acceleration. The radiative acceleration $g_{rad}$ scales with Z$^{1-\alpha}$ \citep[see Eq. 12 and 79 of][]{puls00}, $\alpha$ being the classical CAK parameter. But $g_{rad}$ also depends on $\rho^{-\alpha}$, and since $\rho$ (the density) is directly proportional to \mdot, which in turn scales roughly as Z$^{\frac{1-\alpha}{\alpha}}$ \citep[Eq.\ 87 of ][]{puls00}, all Z effects cancel out. In this context, no metallicity effects are expected if clumping is driven by radiative linear instabilities. Alternatively, \citet{go95} have shown that in the case of multiple scattering, line overlap leads to a reduction of the linear instability growth rate. Presumably, lowering the metallicity would reduce this overlap and produce an enhanced growth rate (S. Owocki, priv. comm.). In that case, clumping might be stronger at low Z. This is however still widely speculative, and future hydrodynamical simulations, including non linear effects, will certainly shed more light on this issue.

%%-------------------------------   evolutionary status   --------------------- 

\subsection{Evolutionary status}
\label{s_evol}

\begin{figure*}
\begin{center}
\begin{minipage}[b]{0.32\linewidth} % A minipage that covers half the page
\centering
\includegraphics[width=6cm]{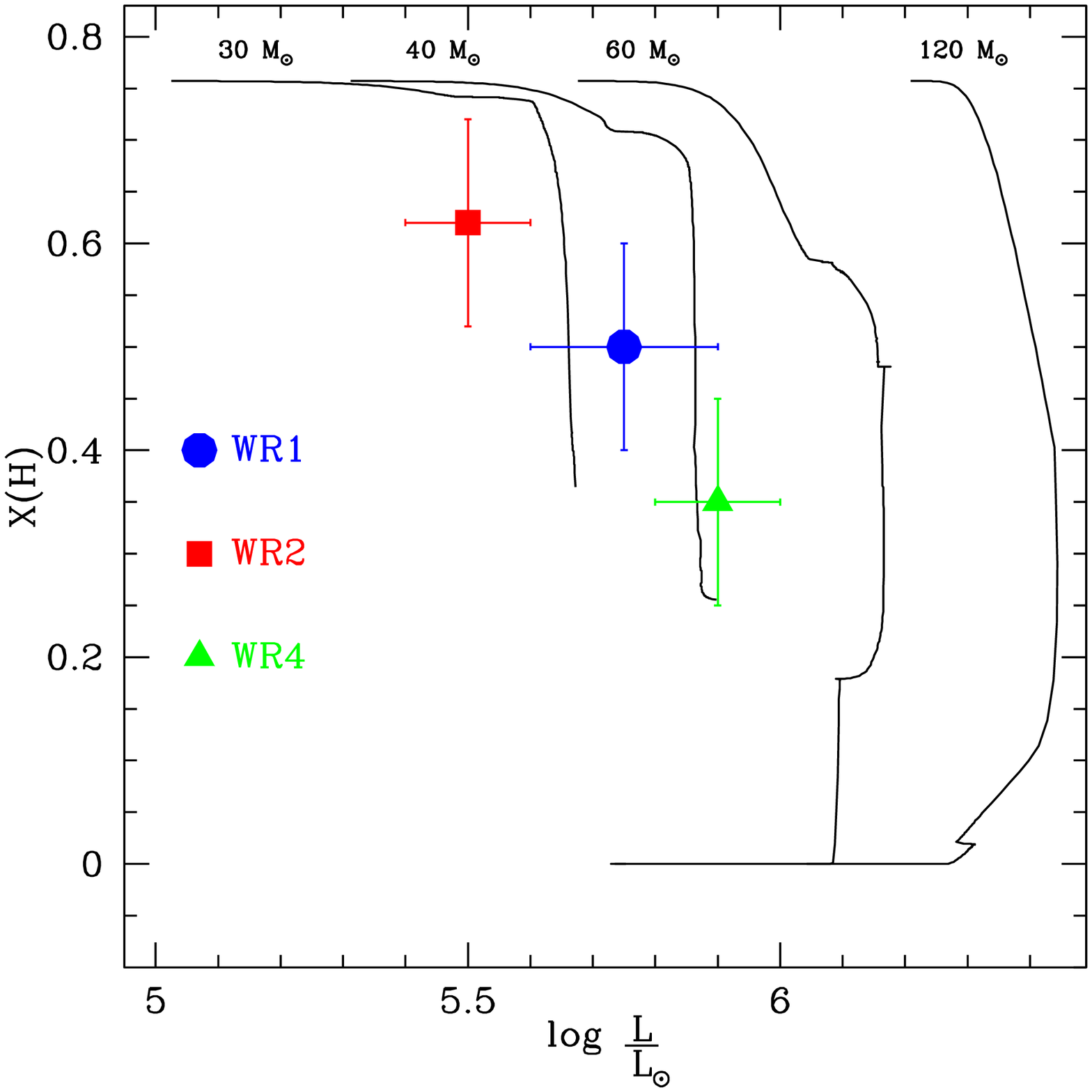}
\end{minipage}
\hspace{0.05cm} % To get a little bit of space between the figures
\begin{minipage}[b]{0.32\linewidth}
\centering
\includegraphics[width=6cm]{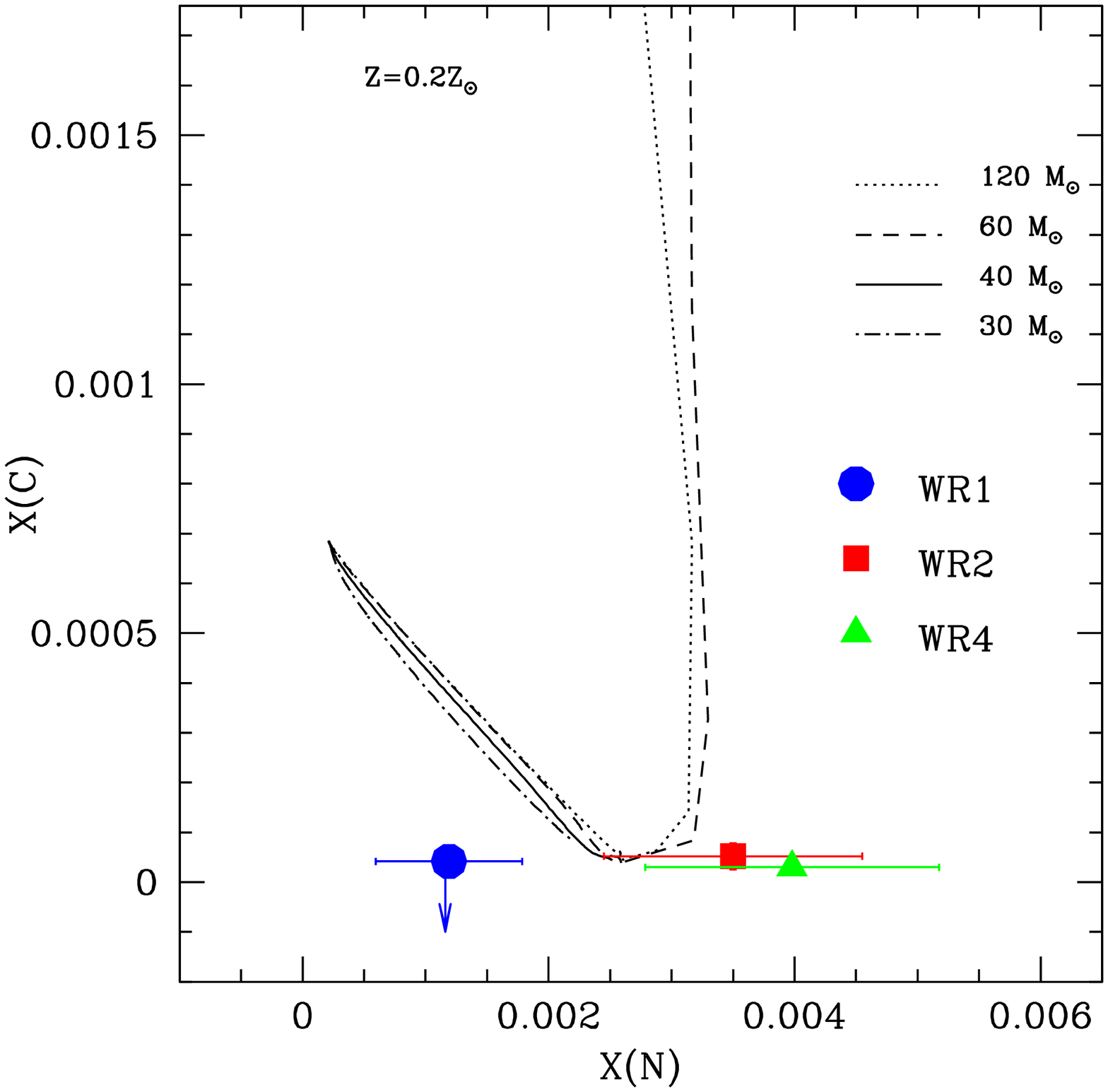}
\end{minipage}
\hspace{0.05cm} % To get a little bit of space between the figures
\begin{minipage}[b]{0.32\linewidth}
\centering
\includegraphics[width=6cm]{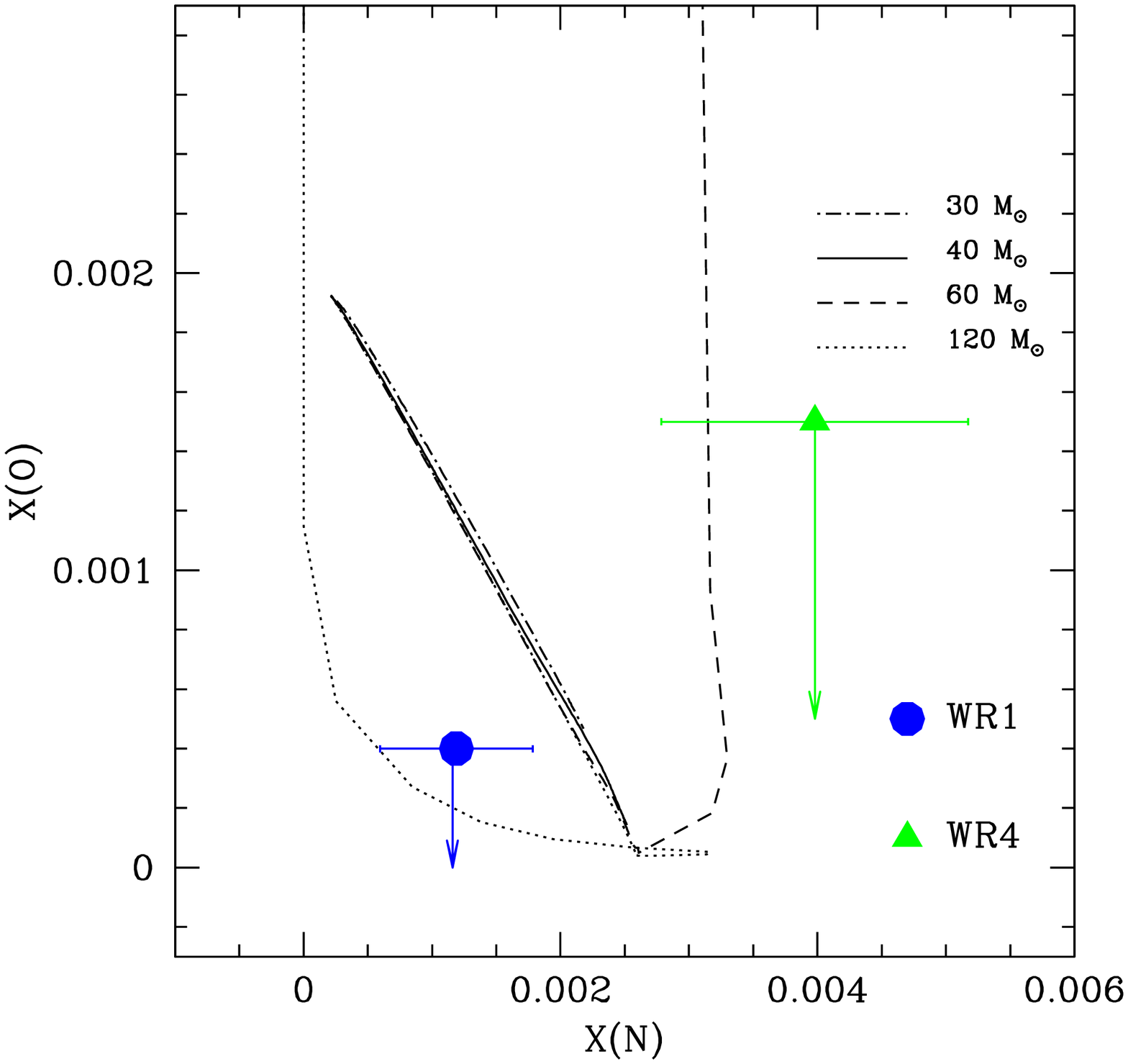}
\end{minipage}
\caption{Evolutionary status from chemical composition. {\it Left:} Hydrogen mass fraction as a function of Luminosity. {\it Middle:} Carbon mass fraction as a function of Nitrogen mass fraction for stars SMC-WR2 and SMC-WR4.  {\it Right}: O mass fraction as a function of N mass fraction from evolutionary tracks (lines) and as derived for star SMC-WR1. The three stars are still rather H rich, but already show signs of CNO processing. Evolutionary tracks are from \cite{mm05} and have an initial rotational velocity of 300 \kms.}\label{hcno}
\end{center}
\end{figure*}

We have shown that the three WNh stars analyzed are slightly He rich, strongly N rich, and C (and O for SMC-WR1 and SMC-WR4) poor. This qualitatively indicates that they are massive stars in an early phase of their post--MS evolution. The large quantity of H still present in their atmosphere shows that they are still H--burning objects, or that they have entered the core He--burning phase very recently. This is confirmed by the position of the stars in a diagram showing the H mass fraction as a function of luminosity (Fig.\ \ref{hcno}, left): one sees that the three stars are located on the early parts of evolutionary tracks, still far away from classical H--free WR stars. In this diagram, all stars can in principle be accounted for by evolutionary tracks.  This was not the case in the earlier study of SMC-WR4 by \citet{paul00}: he found that SMC-WR4 was underluminous by roughly 50$\%$ compared to the closest track of \citet{meynet94}. Crowther concluded that stellar evolutionary tracks were probably missing an important ingredient. This ingredient was stellar rotation, which is included in the evolutionary tracks of \citet{mm05} used in Fig.\ \ref{hcno} (the initial rotational velocity in these tracks is 300 \kms). From this diagram, one can estimate the initial masses of our program stars by interpolating between existing tracks (at $Z = 0.2 Z_{\odot}$). The results are given in the first column of Table \ref{m_estim}: masses range from about 25 \msun\ (SMC-WR2) to about 40 \msun\ (SMC-WR4, SMC-WR1). The errors take into account the uncertainties on \teff\ and \lL. 

The middle panel of Fig.\ \ref{hcno} shows the C versus N mass fraction together with the predictions of \citet{mm05} for various initial masses. The errors on the carbon content are smaller than the symbols. SMC-WR4 and SMC-WR2 are in (or very close to) a phase where X(C) is minimum and X(N) is maximum. SMC-WR1 appears to have a carbon mass fraction consistent with CNO equilibrium, but its N content is a little too low. A similar conclusion is reached if one looks at the right panel of Fig.\ \ref{hcno}: it shows the O versus N mass fraction for stars SMC-WR1 and SMC-WR4. As for carbon, the oxygen content is reduced during the CNO cycle, so that an O underabundance can be expected from evolutionary tracks. The O mass fraction of SMC-WR1 is consistent with the expected minimum, as is the case for carbon. In this figure too the N mass fraction is not as large as expected for the O content. We will see later that the current tracks might not be appropriate for this star. The upper limit on the O content of SMC-WR4 is too high to draw any meaningful conclusion. This limit (not shown) is even higher for star SMC-WR2.

\begin{table}
\begin{center}
\caption{Initial mass estimate (in solar units) from evolutionary diagrams. The second and third column give estimates based respectively on the X(H)--L and HR diagram. The errors take into account uncertainties on \teff\ and \lL. Tracks with $Z = 0.2 Z_{\odot}$ from \citet{mm05} are used.} \label{m_estim}
\begin{tabular}{clrrrrrrrrrr}
\hline
Star    & X(H)--L             & \teff\--L \\
\hline           
        &                     &           \\
1       &  34.4$^{+10.1}_{-7.8}$ & $>$40 \\
        &                     &           \\
2       &  22.8$^{+5.4}_{-5.8}$ & 47.3$^{+7.0}_{-10.6}$ \\
        &                     &           \\
4       &  42.2$^{+6.9}_{-7.8}$ & 64.7$^{+16.0}_{-10.1}$ \\
        &                     &           \\
\hline
\end{tabular}
\end{center}
\end{table}

\begin{figure*}
\begin{center}
\begin{minipage}[b]{0.4\linewidth} % A minipage that covers half the page
\centering
\includegraphics[width=7cm]{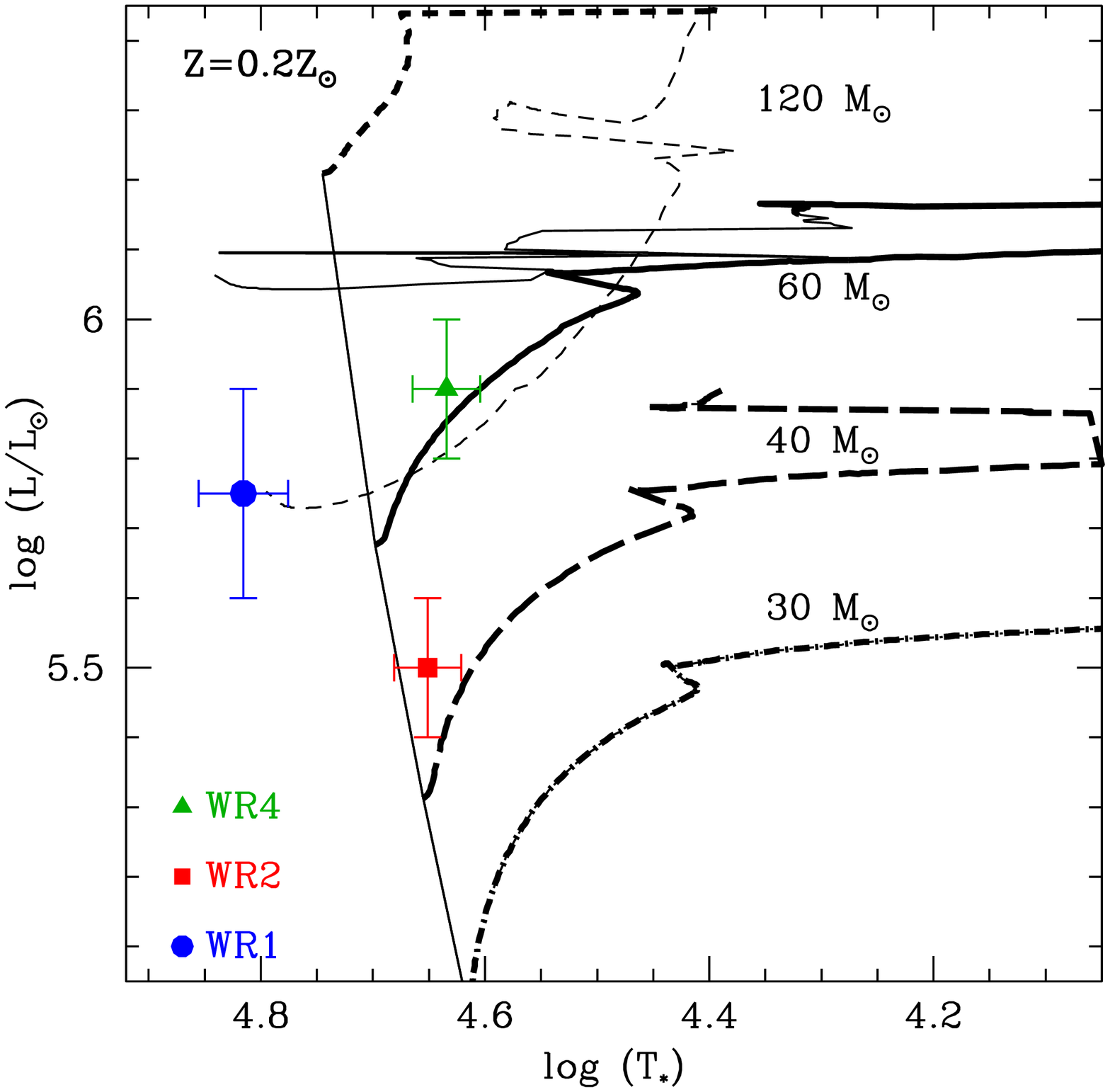}
\end{minipage}
\hspace{0.5cm} % To get a little bit of space between the figures
\begin{minipage}[b]{0.4\linewidth}
\centering
\includegraphics[width=7cm]{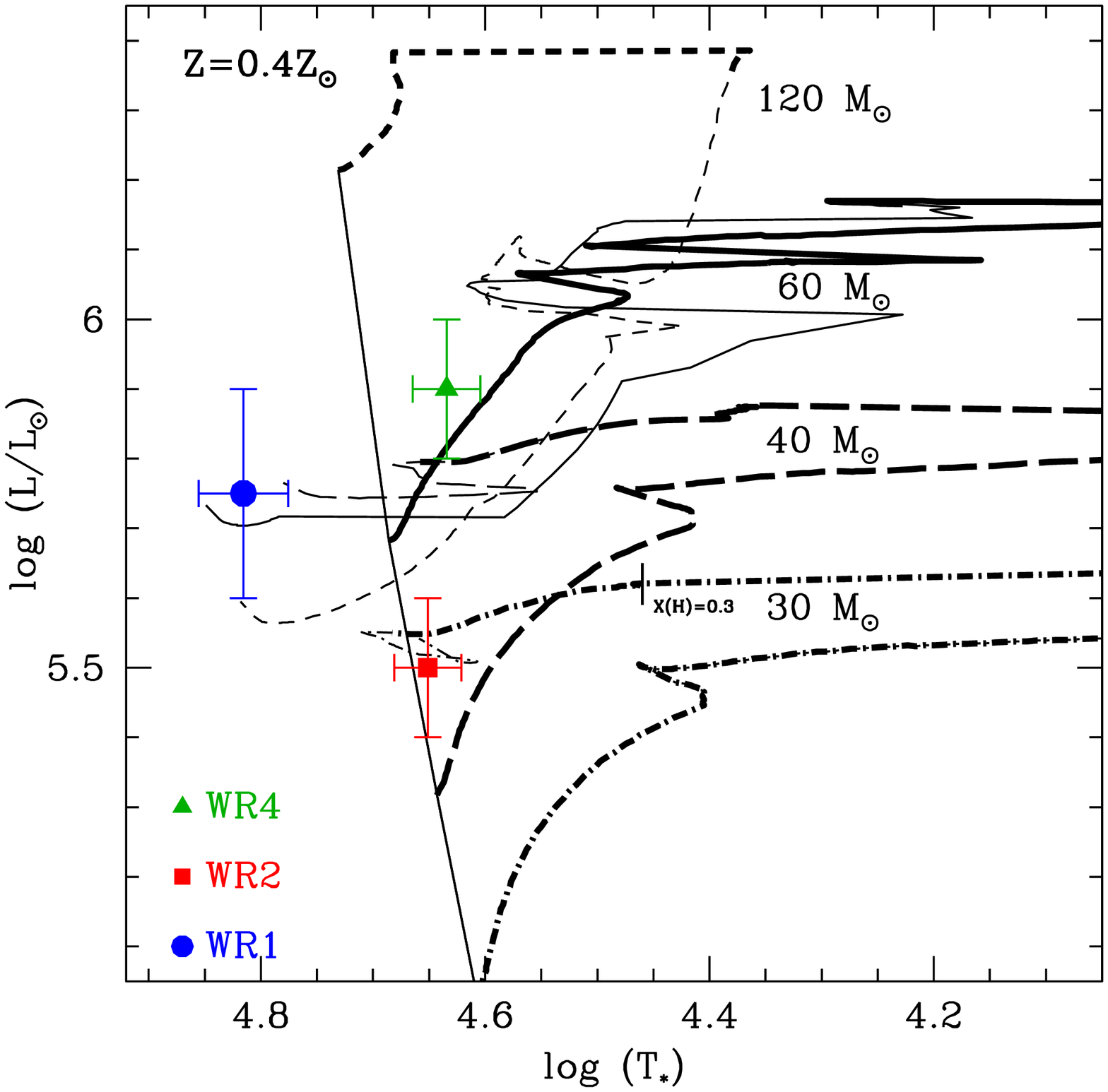}
\end{minipage}
\caption{HR diagram with the three sample stars. Evolutionary tracks are from \cite{mm05}, the left panel being for Z=0.2\zsun, the right one for Z=0.4\zsun. We have highlighted the parts of the tracks corresponding to the H-rich phase, i.e. X(H)$>$20$\%$, using bold lines. On the Z=0.4\zsun\ 30 \msun\ track, we have also noted the position where X(H)=0.3.}\label{hrd}
\end{center}
\end{figure*}

Let us now turn to the HR diagrams shown in Fig.\ \ref{hrd}. The position of the three target stars is indicated on top of the \citet{mm05} evolutionary tracks at Z=0.2 Z$_{\odot}$ (appropriate for the SMC, left panel) and Z=0.4 Z$_{\odot}$ (right panel). We have highlighted in bold font the phase of the tracks in which the H mass fraction is larger than 20$\%$. Since SMC-WR1, SMC-WR2 and SMC-WR4 all have X(H) $>$ 0.3, any comparison between the position of the stars and evolutionary models should rely on the bold lines in either panel. Let us consider each star one by one:

\begin{itemize}

\item SMC-WR4: for this star, we find X(H)=0.35$\pm$0.1. From the left panel of Fig.\ \ref{hrd} (i.e. the one with evolutionary tracks at 0.2 \zsun), one might invoke two scenarios to explain the position of SMC-WR4. In the first one SMC-WR4 is a relatively unevolved star, still close to the main sequence and with an initial mass slightly above 60 \msun. In that case the initial mass can be estimated from the uncertainties on \teff\ and \lL, as given in the second column of Table \ref{m_estim}. The result is somewhat surprising, since it is much larger than the estimate based on the X(H)--L diagram, with no overlap within the errors (42.2$^{+6.9}_{-7.8}$ vs 64.7$^{+16.0}_{-10.1}$). Besides, if SMC-WR4 were indeed at an early post--main sequence phase of its evolution, its mass loss rate would be around $10^{-5.7}$ \myr\ according to \citet{vink01}. Instead, we find a value one order of magnitude larger (once corrected for clumping). Thus it appears unlikely that SMC-WR4 is just evolving off the main sequence.

Another possibility is that SMC-WR4 is a star with an initial mass of 40--50 \msun\ and evolving blue-ward from a supergiant phase. Indeed, the $Z = 0.2 Z_{\odot}$ 40 \msun\ track ends a little before reaching the position of SMC-WR4. However, the same track at Z=0.4 \zsun\ passes through the position of SMC-WR4 while still having X(H)$>$0.2 (see right panel of Fig.\ \ref{hrd}). Hence assuming a metal content slightly larger than 0.2 \zsun\ can explain the characteristics of SMC-WR4. In that case, the 40 \msun\ track would explain the position of SMC-WR4 in both the HR and the X(H)--L diagram (the initial mass derived from the latter diagram is very similar at 0.2 and 0.4 \zsun\ for SMC-WR4). Note that other effects such as a slightly different initial rotation rate compared to the values used in the evolutionary calculations of \citet{mm03}, or different mass loss rates (uncertainties being large due to clumping) could also modify the tracks at Z=0.2 \zsun\ and explain the position of SMC-WR4. Given this, we tentatively conclude that SMC-WR4 results from rather normal evolution. \\

\item SMC-WR2: the hydrogen mass fraction is the largest of the three stars, reaching 0.62$\pm$ 0.1. The situation is different from SMC-WR4. There is only one possibility to explain both the effective temperature/luminosity and H content of SMC-WR2: it must be a star just evolving off the main sequence. Indeed, none of the evolutionary tracks coming back from the cool part of the HR diagram and with low enough luminosities can account for the large H mass fraction of SMC-WR2. The 30\msun\ track at Z=0.4\zsun\ has a too low H content when reaching the position of SMC-WR2 (see the tick on the track at X(H)=0.3). The star has to be in its early evolution. If we accept this, one can estimate the initial mass of the star by interpolation between the red-ward evolving tracks. The results (given in the third column of Fig.\ \ref{m_estim}) indicate M=47.3$^{+7.0}_{-10.6}$ \msun. This is significantly larger than the estimate from the X(H)--L diagram (22.8$^{+5.8}_{-5.4}$ \msun). Such a difference is surprising, since it is based on the same set of tracks. We will discuss possible solutions to this puzzle in Sect.\ \ref{s_nat}. \\

\item SMC-WR1: The situation is even more intriguing for star SMC-WR1. We find convincing evidence that SMC-WR1 is still H--rich, with X(H)=0.50$\pm$0.1. But a look at Fig.\ \ref{hrd} shows that SMC-WR1 is definitely located to the left of the H--burning main sequence, at a temperature of $\sim$65000 K and \lL\ of 5.75. The only evolutionary tracks reaching this part of the diagram have initial masses larger than $\sim$ 50 \msun. But this is only at the very end of the star's evolution, when it is H and N free. The properties of SMC-WR1 are thus extremely disturbing, classical stellar evolution being unable to explain them.  

\end{itemize}

From this analysis, one can conclude that one star (SMC-WR4) seems to have followed a normal evolution. On the contrary, the two remaining stars (SMC-WR1 and SMC-WR2) have probably been affected by uncommon physical processes. We tentatively identify some of them in Sect.\ \ref{s_nat}. 

Before concluding this section, we briefly note that mass estimates are important clues to the nature of the sample stars. One can add another such estimate assuming that the stars have Eddington luminosities. In that case one obtains lower limits to the current masses of the three WNh stars. These values are respectively 9.8, 7.8 and 19.6 \msun\ for SMC-WR1, SMC-WR2 and SMC-WR4 (assuming the only source of opacities is electron scattering). In practice, our modelling indicate that the maximum Eddington factors in their atmosphere are of the order 0.7 and that the total radiative acceleration is 1.2--1.5 times the acceleration due to electron scattering, implying current masses of the order 10-35 \msun. Of course, these values have to be taken as rough estimates since they depend on the exact hydrodynamical structure which still suffers from uncertainties (see Sect.\ \ref{s_models}).

%%-------------------------------   nature of the stars   --------------------- 

\subsection{Evidence for evolution with fast rotation.}
\label{s_nat}

The previous section has shown that classical stellar evolution was able to reproduce the observed properties of SMC-WR4 if its metallicity is a little larger than 0.2\zsun. However, it was shown that the current set of evolutionary tracks failed to explain SMC-WR1 and SMC-WR2.

Before considering SMC-WR1 and SMC-WR2, let us first reiterate our conclusions regarding SMC-WR4. We showed that both its position in the HR diagram and its surface abundances were compatible with a star of initial mass 40--50 \msun\ in or about to enter the core He burning phase. This is rather similar to the conclusion of \citet{foel03a} who argued that SMC-WR4 most likely resulted from standard evolution of a 70--85 \msun\ star. We stress once more that according to this scenario, SMC-WR4 is coming back from the red part of the HR diagram. Consequently, it is not a core H burning object. This is important in the context of the understanding of the properties of H rich WN stars. Recently, \citet{arches} analyzed 16 WN7--9h stars in the Arches cluster and showed that they were rather young and very massive stars still in the core H burning phase. The WN6h stars in the core of NGC3603 are probably similar objects \citep{cd98}. The initial mass estimates for these objects in the Arches and NGC3603 cluster range between 60 and 130 \msun\ (see \citet{schnurr08} for a recent keplerian mass estimate of NGC3603/A1). Hence, it seems that H--rich WN stars come in different flavors. They can be very massive objects in an early phase of evolution, with ages not larger than 2--3 Myr; but they can also result from less massive objects and be more evolved. In the latter category, one should mention star WR3 in the outer part of the Galaxy. \citet{mar04} showed that it was a H--rich object resulting from the evolution of an initially 40--50 \msun\ object. Since lower metallicity is encountered in the outer regions of the Galaxy, star WR3 might be more similar to the SMC stars analyzed in the present study.

Coming back to SMC-WR1 and SMC-WR2, there are a few possibilities to explain the puzzling results of Sect.\ \ref{s_evol}. The first one is binarity. \citet{well99} showed that mass transfer during binary evolution could dramatically change the evolutionary sequence of both components of the system. In particular, their Fig.\ 1 shows that the primary might evolve almost vertically just after the mass transfer. Consequently, if such a star was analyzed by means of classical single star tracks, its mass would be overestimated. This could explain the inconsistencies encountered for SMC-WR2: its mass derived from the HR diagram is larger than that derived from the X(H)--L diagram. Besides, mass transfer would affect the surface abundances, so that the derived H mass fraction might be different from that of single evolutionary sequences. Hence, mass estimates from the X(H)--L diagram would be affected too. Binarity might thus be an alternative to explain the properties of SMC-WR2. However, \citet{foel03a} thoroughly studied the radial velocities and photometry of all the SMC stars and showed that SMC-WR2 was not varying. They concluded that it was most likely a single star, or maybe a long period binary. But in any case it was not a short period binary prone to experience the kind of evolution described above. The same conclusion was reached for SMC-WR1. Periodic changes of the shape of emission profiles may also reveal the close-binary nature of a massive star.
The line-profile variability of SMC-WR1 and SMC-WR2 closely resembles the behavior of emission lines in the presumably single Galactic W-R stars (Paper I), where the stochastic, seemingly random variability is induced by numerous wind-embedded clumps. Hence, we conclude that practically all the variability patterns seen in SMC-WR1 and SMC-WR2  are caused by  stochastic appearance of small-scale overdensities in the winds of the SMC W-R stars.

Binarity thus appears unlikely to be the explanation for the properties of SMC-WR1 and SMC-WR2. Alternatively, there is growing evidence that homogeneous evolution might be at work in at least a fraction of massive stars in the Magellanic Clouds. \citet{jc03} argued that star MPG 355 in NGC 346 could have the same age as the other cluster members if it was evolving homogeneously. \citet{walborn04} explained the position of highly N enriched Magellanic Cloud O2 stars in the HR diagram by large mixing leading to homogeneous evolution. Finally, \citet{mokiem07} explained the correlation of large mass discrepancies \footnote{The ``mass discrepancy'' problem refers to the larger evolutionary masses derived from HR diagrams compared to spectroscopic masses derived from gravities through atmosphere modelling \citep[see][]{her92}.} with He enrichment by the fact that He--rich stars might evolve homogeneously, so that using normal tracks would overestimate their evolutionary mass. 

Homogeneous evolution was shown to occur in rapidly rotating stars half a century ago by \citet{sch58}. More recently, \citet{maeder87} and \citet{langer92} ran evolutionary models confirming that fully mixed stars were evolving blue-ward in the HR diagram. The conditions under which such homogeneous chemical evolution occurs depend on several parameters. The basic constraint is that the timescale for rotationally induced mixing should be shorter than the nuclear timescale. In practice, this means that the stars should keep a high rotational rate during their evolution. A key ingredient to keep the high rotational rate is the strength of the stellar wind. Larger mass loss rates will generally lead to larger angular momentum removal, which in turn will slow down the rotation. As radiatively driven winds are weaker at lower metallicity \citep[e.g.][]{puls00}, changes in the surface abundances are expected to be more common at (very) low metallicity. But the downward revision of mass loss rates of massive stars due to clumping \citep[e.g.][]{paul02,jc05} could make homogeneous evolution more common at higher metallicity than previously thought. Besides, mass loss anisotropy implied by fast rotation might affect the details of the subsequent evolution \citep{mm07}. In addition, magnetic fields can modify the way rotation changes during the lifetime of the star, leading to solid body rotation \citep{mm03}. From this, one sees that homogeneous chemical evolution will occur under a combination of effects. A systematic study of these conditions is still lacking, partly because of the uncertainties in the mass loss rates of massive stars, especially Wolf-Rayet stars. 

\begin{figure*}
\begin{center}
\begin{minipage}[b]{0.4\linewidth} % A minipage that covers half the page
\centering
\includegraphics[width=7cm]{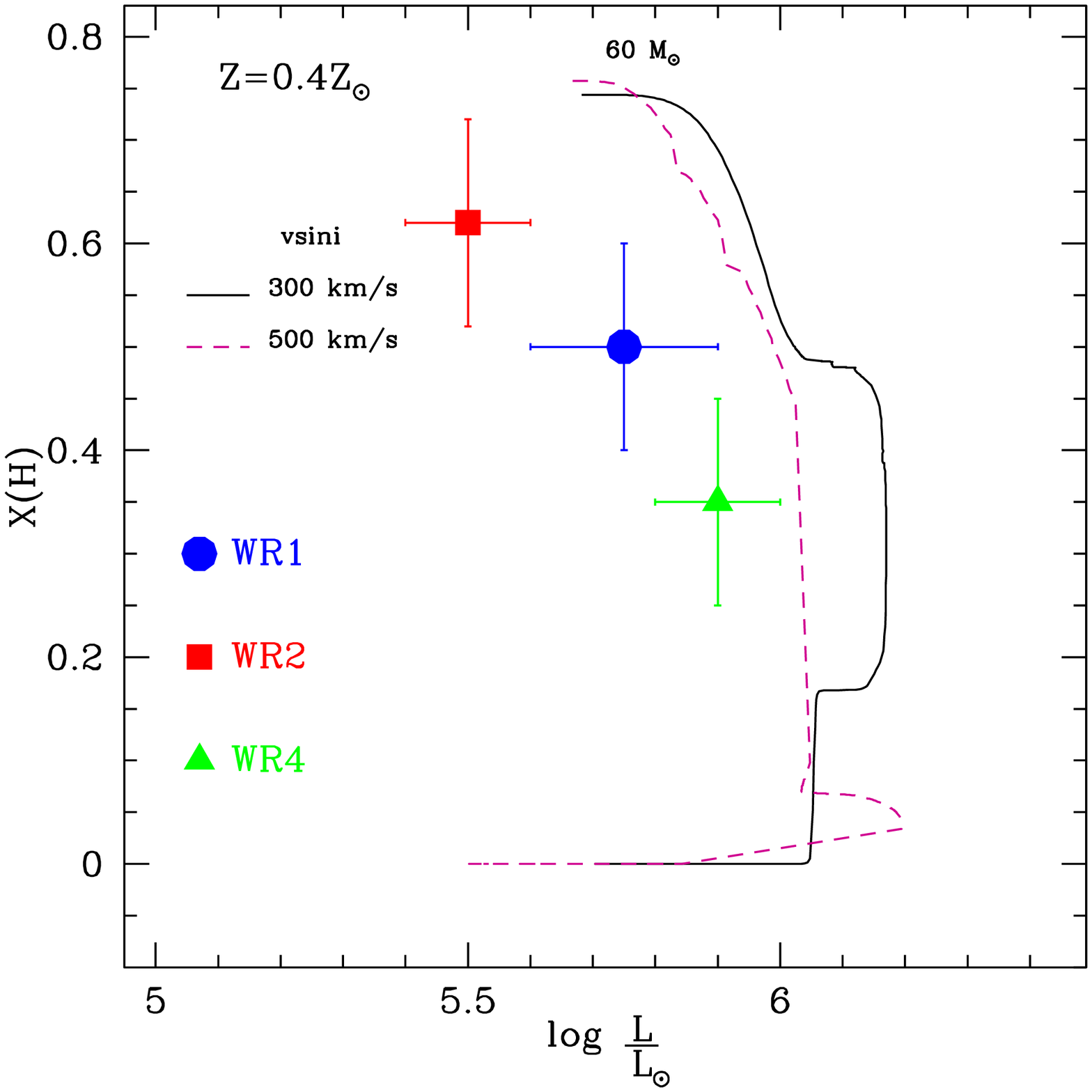}
\end{minipage}
\hspace{0.5cm} % To get a little bit of space between the figures
\begin{minipage}[b]{0.4\linewidth}
\centering
\includegraphics[width=7cm]{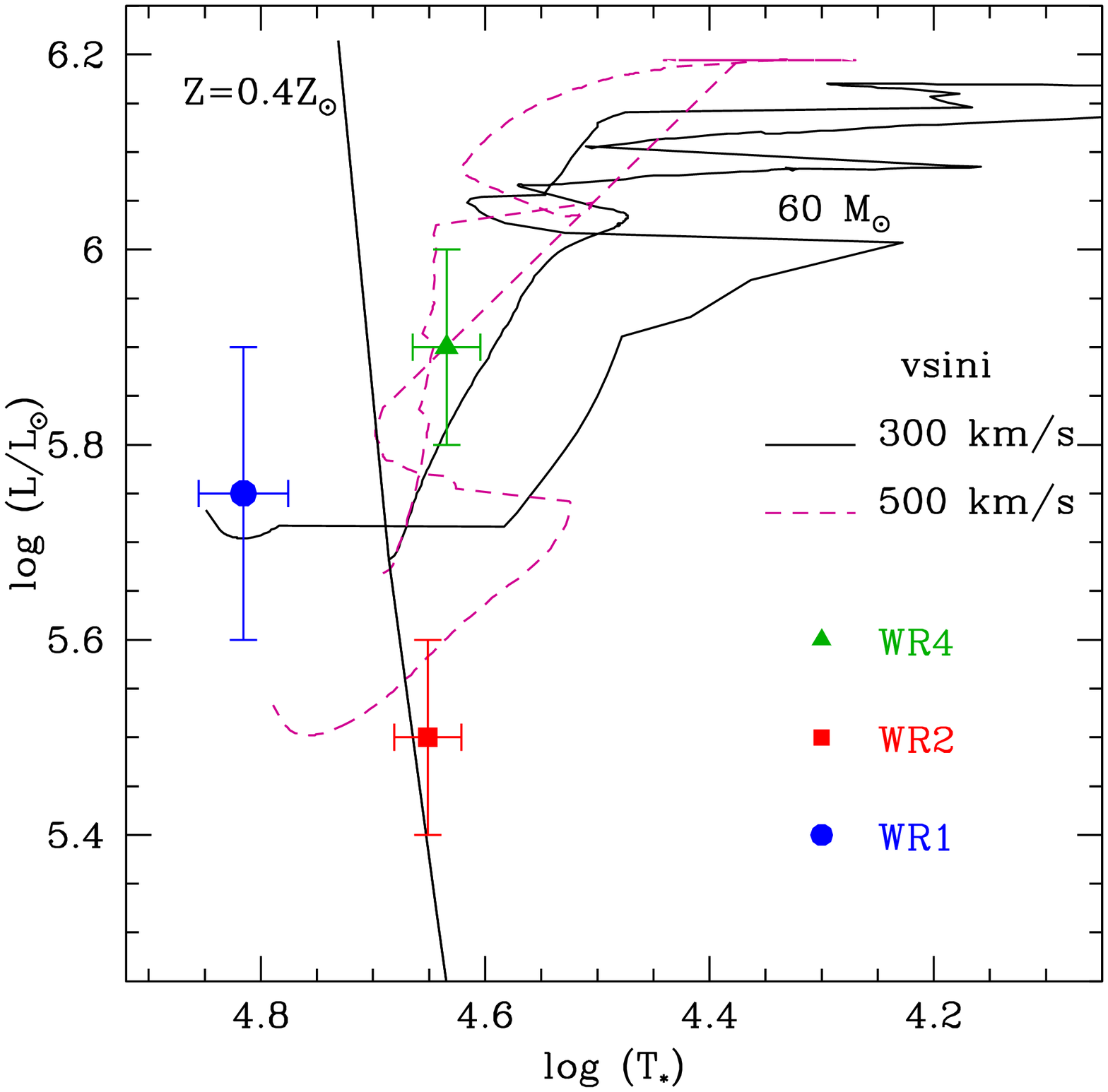}
\end{minipage}
\caption{Illustration of the effect of rotation on the X(H)--L diagram (left panel) and on the HR diagram (right panel). The masses estimated using different rotational velocities will be different in both diagrams. Evolutionary tracks from \citet{mm05}. }\label{effect_rot}
\end{center}
\end{figure*}

Nonetheless, various works have mentioned the existence of homogeneous evolution. \citet{mm00a} argue that stars initially more massive than 40 \msun\ and with $\frac{\Omega}{\Omega_{crit}} > 0.5$ -- $\Omega$ ($\Omega_{crit}$) being the (critical) rotation rate -- will evolve homogeneously. The 60 \msun\ evolutionary track of \citet{mm05} with an initial \vsini\ of 500 \kms, as well as the fast rotating models of \citet{mm07} (which include the effects of a magnetic field) confirm this trend. \citet{yl05} also produce blue-ward evolving stars in their fast rotating models at low metallicity. As mentioned above, a complete set of evolutionary tracks for various metallicities and rotational rates does not exist at present. Only a few tracks are available. In Fig.\ \ref{effect_rot}, we plot the evolutionary tracks of a 60 \msun\ star from \citet{mm05} for both \vsini\ = 300 \kms\ (solid line) and 500 \kms\ (dashed line). The latter corresponds to a borderline case where the evolution is almost homogeneous, the star evolving at almost constant \teff\ in the first phases. In the X(H)--L diagram, one sees that for fast rotating stars, the track is shifted to lower luminosities. In practice, this means that the mass one would derived from a set of tracks with \vsini\ = 500 \kms\ would be larger than for the case where \vsini\ = 300 \kms. On the contrary, fast rotating stars are on average more luminous in the HR diagram (due to the effects summarized above). Hence, lower masses are derived for larger \vsini\ tracks. This is interesting in the context of the mass problem reported for star SMC-WR2: the large difference between the masses derived from the X(H)--L and HR diagram can, at least qualitatively, be reduced if one uses faster rotating tracks. This strengthens the idea that SMC-WR2 might be a rapidly rotating star, close to homogeneous evolution. It is worth mentioning that using the two methods to estimate masses we have applied in this paper, one might have an indirect way of determining the initial rotational velocity of Wolf-Rayet stars: a set of rotating tracks should be able to give consistent mass estimates in the X(H)--L and HR diagram. If not, the initial rotational velocity is probably not correct.

The case of SMC-WR1 is more extreme: it is located on the left side of the main sequence, but still contains a significant fraction of hydrogen. Inspection of Fig.\ 1 of \citet{mm07} reveals that homogeneous evolution can reproduce these features. In this figure, SMC-WR1 would lie close to  to the anisotropic wind track ($\log T_{\rm eff} = 4.8, \log \frac{L}{L_{\odot}}=5.75$) -- although slightly below due to the lower mass of the star compared to the model --, where the core (and consequently the surface) still contains hydrogen. Qualitatively, SMC-WR1 is thus likely an example of a WR star proceeding from homogeneous evolution of a massive star. Star WR3 in the Galaxy is probably a twin of SMC-WR1. \citet{mar04} showed that it was H--rich and very similar spectroscopically to early WN stars in the SMC. Given our results, it is likely that WR3 also results from evolution with high rotation.

As stated above, no systematic study of the occurrence of homogeneous chemical evolution with metallicity, magnetic field and rotation velocity exists. However, a key ingredient is the high initial rotation rate. One might wonder if initial rotational velocities as large as 500 \kms\ can be reached in the SMC. Recent studies of the projected rotational velocities of B stars in the SMC were carried out by \citet{martayan07} and \citet{hunter08}. Both works show that the \vsini\ distribution is shifted towards larger values in the SMC compared to the Galaxy, with averages around 155--175 \kms. According to \citet{hunter08}, about 10$\%$ of the SMC stars have \vsini\ larger than 300 \kms. Considering that these values are lower limits on the true surface velocities (the inclinations being unknown), stars rotating at 500 \kms\ might exist in the SMC. We thus argue that SMC-WR1 is most likely the result of homogeneous evolution of a massive star. 

\begin{figure}[]
\centering
\includegraphics[width=9cm]{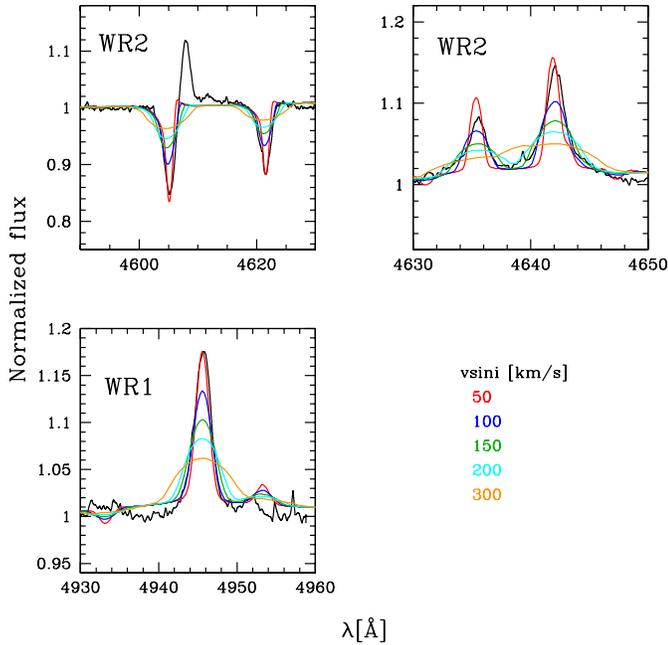}
\caption{Estimate of the current projected rotational velocity of stars SMC-WR1 and SMC-WR2. The black solid lines are the observed spectra of SMC-WR2 (top panels) and SMC-WR1 (bottom panel). Colored lines are synthetic spectra convolved with rotational profiles with different \vsini. Values larger than 50 \kms\ are excluded. }\label{vsini_wr12}
\end{figure}

We have seen that one could get insight into the initial rotational velocity of our target stars by using both the X(H)--L and HR diagrams. In Fig.\ \ref{vsini_wr12} we constrain the current projected rotational velocities of SMC-WR1 and SMC-WR2 from the shape of narrow absorption/emission lines. One clearly sees that values much larger than 50 \kms\ are excluded, the line profiles becoming much too broad. These values are lower limits on the true rotational velocities. However, it is unlikely that we see both stars pole--on, so that SMC-WR1 and SMC-WR2 probably have equatorial velocities $\lesssim$ 100 \kms. Normal evolutionary tracks with rotation predict equatorial velocities between 0 and 200 \kms\ for 40--60 \msun\ stars in the WR phase \citep{mm05}. On the other hand, the homogeneous tracks of \citet{mm07} predict velocities between 100 and 400 \kms\ in the case of isotropic winds, and around 600 \kms\ for anisotropic winds. The latter predictions (anisotropic winds) are supposed to be more realistic. Since we observe rather modest rotational velocities, and given the strong evidence for homogeneous evolution -- and thus fast initial rotation -- at least for star SMC-WR1, one might argue that braking was more efficient than predicted for these stars.

In the context of GRB formation, fast rotation is the key ingredient (see \citet{wb06} for a review). The other important condition is that the star explodes while it has lost all its hydrogen envelope. Here, we have shown that SMC-WR1 has most likely evolved with a high rotation rate until recently. The fact that it still contains a large H mass fraction might argue against its potential explosion as a GRB. However, according to the model predictions of \citet{mm07}, a star following homogeneous evolution will end its life as a H--free WR star (of the class WC or WO). \citet{hirschi05} argue that GRBs can exist at SMC metallicities and that they should have WO stars as progenitors. Similarly, \citet{yoon06} predict a metallicity threshold of about Z=0.004 for the formation of long--soft GRBs (the exact value depending mainly on the adopted mass loss rates). In their scenario, stars ending their lives as GRBs follow quasi chemically homogeneous evolution. The properties of SMC-WR1 are very reminiscent of this scenario. However, the questions raised in the the previous paragraph prevent us from concluding firmly that SMC-WR1 is on its way to become a GRB progenitor.

%%#####################################################################
%%-------------------------------   Conclusion   ----------------------- 

\section{Conclusion}
\label{s_conc}

We have conducted a detailed spectroscopic analysis of three WNh stars in the SMC. Using atmosphere models computed with the code CMFGEN, we have determined the stellar and wind parameters. The key results can be summarized as follows:

\begin{itemize}

\item[$\bullet$] the quantitative analysis confirms the spectral classification in the sense that all three stars still contain large amounts of hydrogen in their atmosphere. At the same time, they show clear signs of CNO processing, with C (and O for SMC-WR1) depletion, and N enrichment. SMC-WR1 is a hotter object than the SMC-WR2 and SMC-WR4, as suggested by its earlier spectral type.

\item[$\bullet$] SMC-WR4 can be explained by normal evolutionary tracks including rotation if its global metallicity is slightly above 1/5$^{th}$ solar. In that case, it is most likely a star with an initial mass of $40-50$ \msun\ in or very close to the core He burning phase.

\item[$\bullet$] SMC-WR2 and, more dramatically, SMC-WR1 cannot be accounted for by moderately rotating (i.e. 300 \kms) evolutionary tracks. Their position in both the HR and X(H)--L diagrams can be reproduced only if the stars rotate initially very fast ($>$ 500 \kms). In the case of SMC-WR1, only homogeneous evolution can account for the position of the star to the left of the main sequence and the high H mass fraction in its atmosphere. 

\item[$\bullet$] the clumping factor $f_{\infty}$ of star SMC-WR4 is quantitatively constrained to be 0.15$\pm$0.05. Within this uncertainty, no difference is found with Galactic WR stars, confirming the dynamical results of paper I. 

\end{itemize}

The present results are based on a limited number of stars. In a subsequent publication, we will present a study of the remaining Wolf-Rayet population of the SMC. This will provide a wider view of the stellar properties of these objects. In particular, we will be able to estimate the fraction of stars potentially following evolution governed by fast rotation. This will be useful in the context of Gamma-Ray Burst formation, thought to be due to very rapidly rotating WR stars.

A more systematic study of the clumping properties of SMC WNh stars is also required to confirm that clumping does not seem to depend on metallicity. This is crucial to understand the physics of inhomogeneity formation in massive--stars winds and, more generally, the mechanisms of mass loss. Our future study will tackle these questions too.

%%#####################################################################
\begin{acknowledgements}
We thank the referee, Paul Crowther, for useful comments which helped to improve the spectroscopic analysis and for sharing near-infrared spectra. We thank Stan Owocki for illuminating discussions about radiative instabilities. FM thanks the CINES for allocation of computing time. AFJM thanks NSERC (Canada) and FQRNT (Quebec) for financial aid. JCB acknowledges support from the French National Research Agency (ANR) through program number ANR-06-BLAN-0105.
\end{acknowledgements}

%%#####################################################################
\bibliography{biblio.bib}

%%#####################################################################
\begin{appendix}
\section{Best fits}
In this Section, we gather the figures showing the comparison of our best fit models with the observed spectra of our program stars.

\begin{figure*}[]
\centering
\includegraphics[width=19cm,height=25cm]{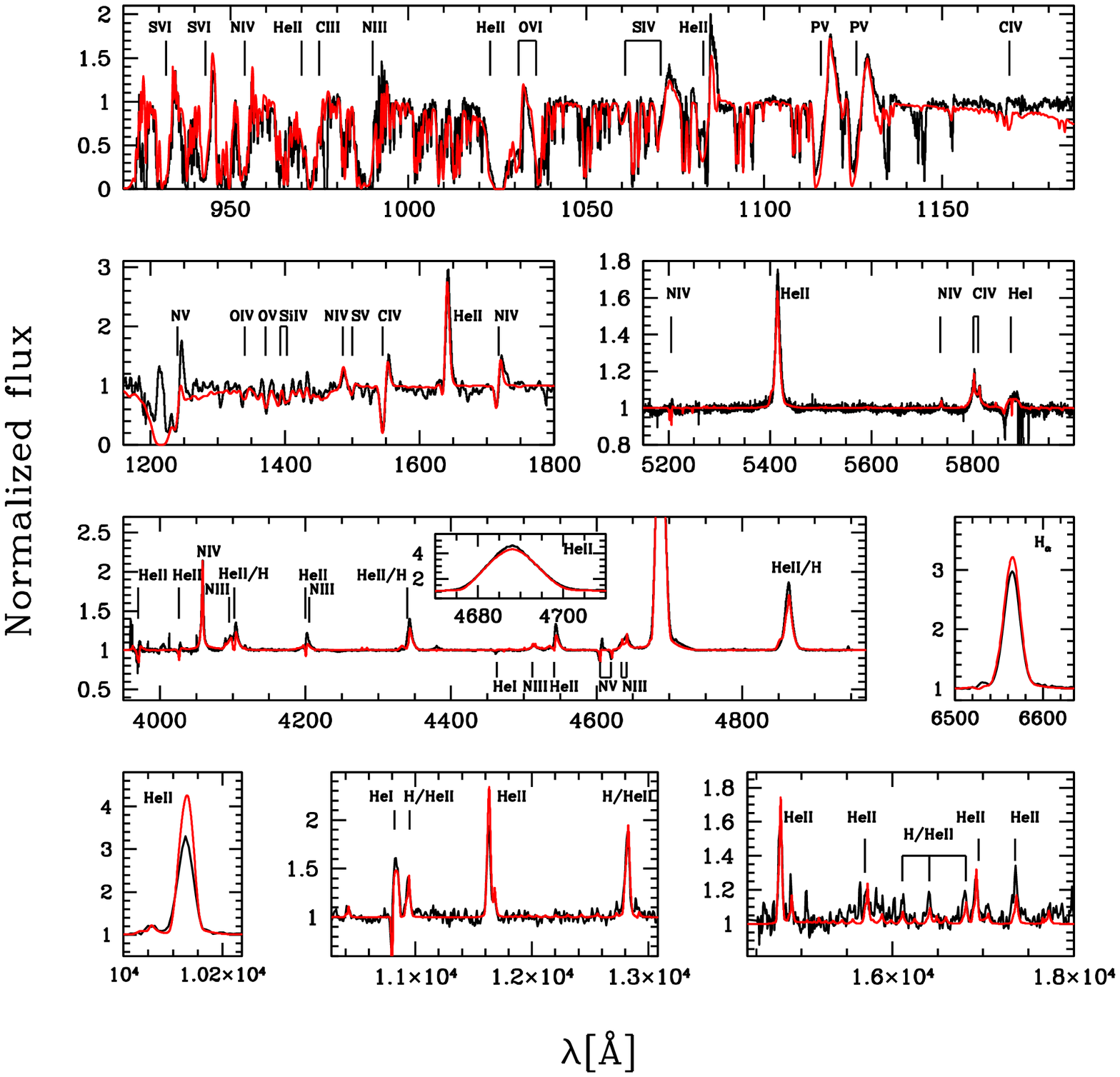}
\caption{Best fit (red solid line) of the observed spectrum (black solid line) of star SMC-WR4. The main stellar lines are indicated.}\label{fit_wr4}
\end{figure*}

\begin{figure*}[]
\centering
\includegraphics[width=19cm,height=25cm]{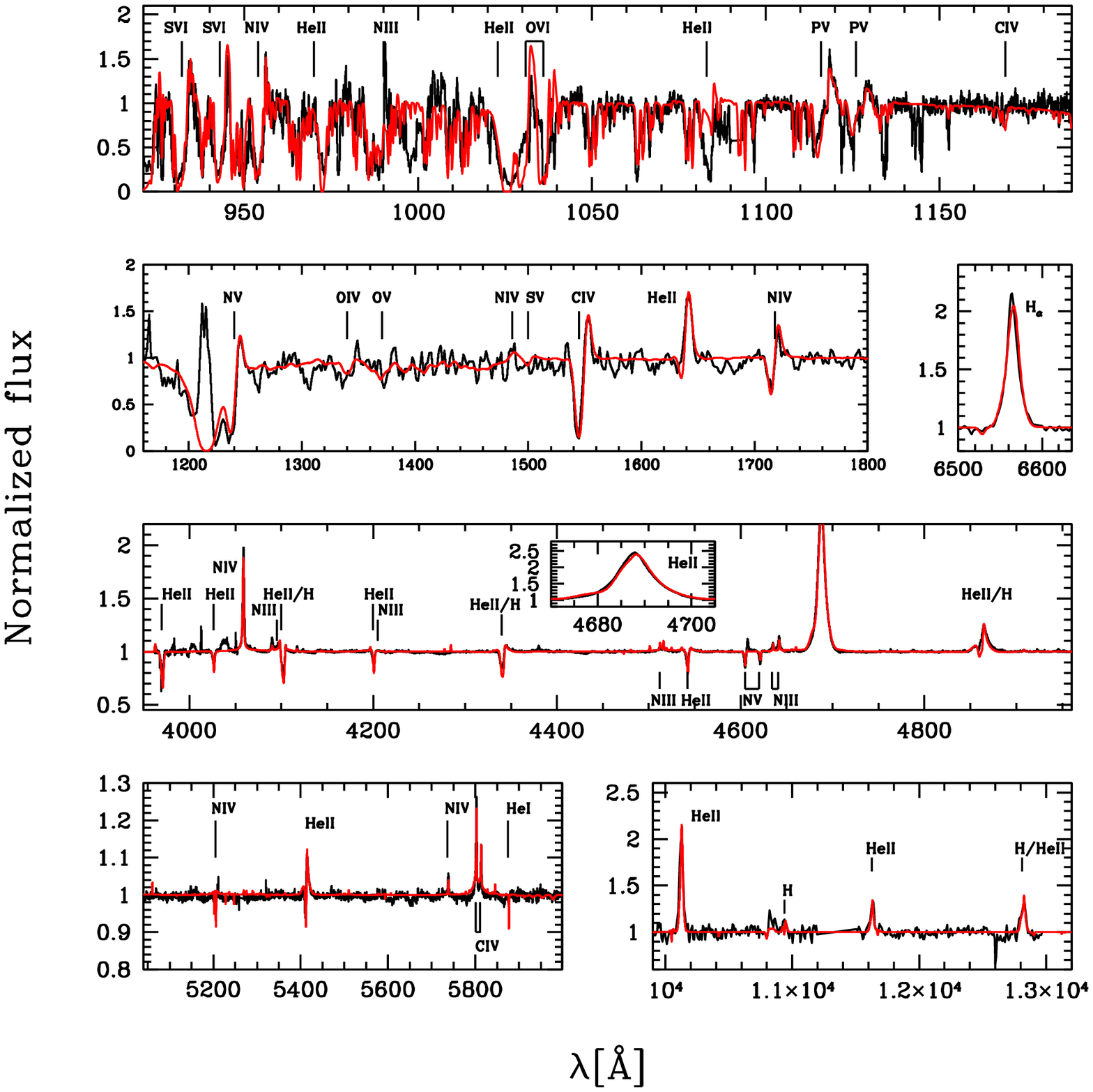}
\caption{Best fit (red solid line) of the observed spectrum (black solid line) of star SMC-WR2. The main stellar lines are indicated.}\label{fit_wr2}
\end{figure*}

\begin{figure*}[]
\centering
\includegraphics[width=19cm,height=25cm]{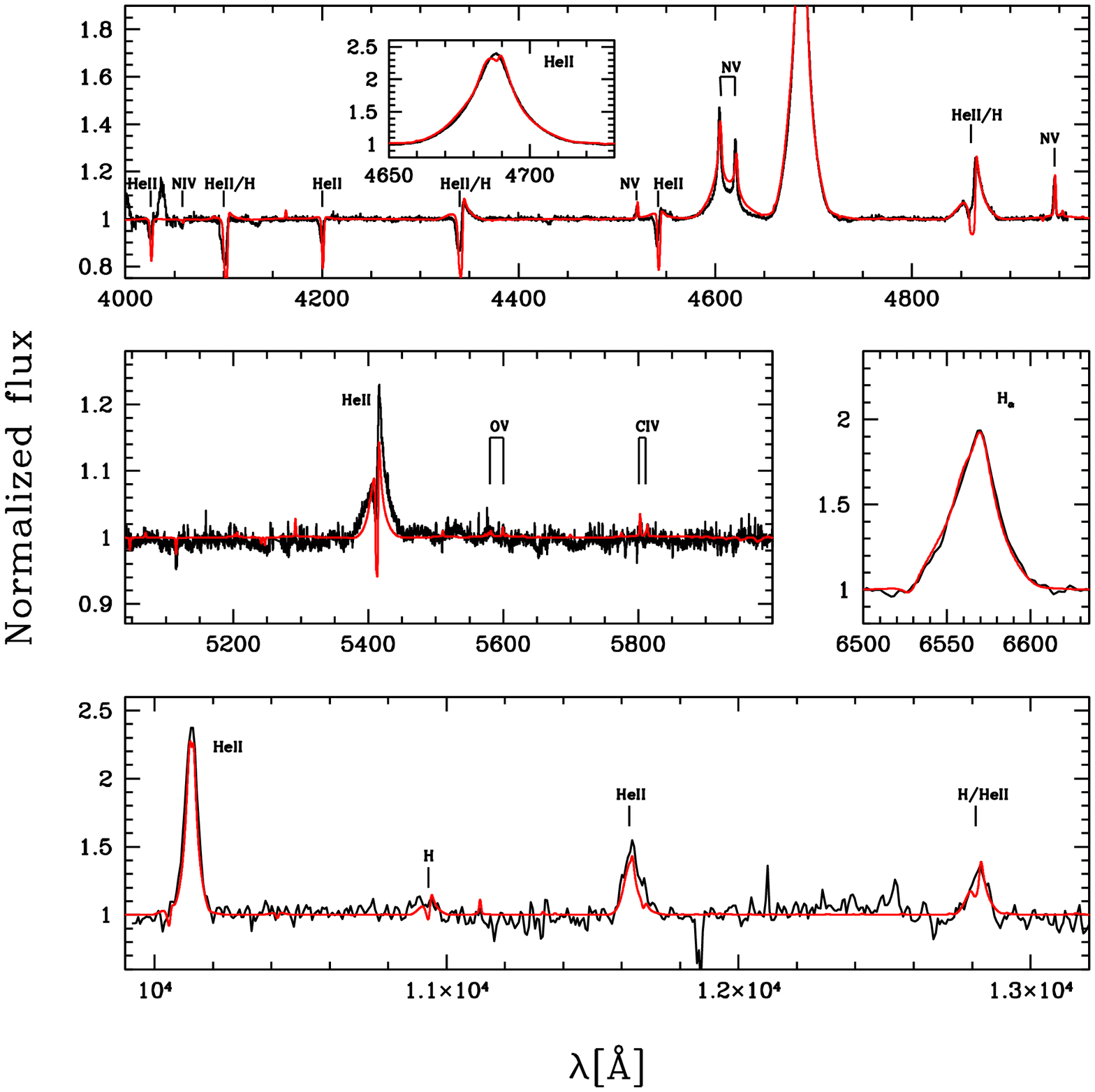}
\caption{Best fit (red solid line) of the observed spectrum (black solid line) of star SMC-WR1. The main stellar lines are indicated.}\label{fit_wr1}
\end{figure*}

\end{appendix}

\end{document}